\documentclass{aa}  
 \usepackage{comment} 
\usepackage{colortbl}
\usepackage{graphicx}
\usepackage{txfonts}
\usepackage[]{caption}
\usepackage{subfigure}
\usepackage{multirow}
\definecolor{pink}{rgb}{1.0, 0, 0.8}
\usepackage{tabularx}
 \def\fdg{\hbox{$.\!\!^\circ$}}  
\begin{document} 
\title{Apsidal motion in massive eccentric binaries: \\The case of CPD-41$^\circ$\,7742, and HD\,152218 revisited}
\author{S.\ Rosu\inst{1}\fnmsep\thanks{Research Fellow F.R.S.-FNRS (Belgium)} \and G.\ Rauw\inst{1} \and Y. Naz\'e\inst{1}\fnmsep\thanks{Senior Research Associate F.R.S.-FNRS (Belgium)}  \and E. Gosset\inst{1}\fnmsep\thanks{Research Director F.R.S.-FNRS (Belgium)}  \and C. Sterken\inst{2}}
\mail{sophie.rosu@uliege.be}
\institute{Space sciences, Technologies and Astrophysics Research (STAR) Institute, Universit\'e de Li\`ege, All\'ee du 6 Ao\^ut, 19c, B\^at B5c, 4000 Li\`ege, Belgium \and Physics Department, Vrije Universiteit Brussel, Pleinlaan 2, 1050 Brussels, Belgium}
\date{}

\abstract{This paper is part of a study of the apsidal motion in close eccentric massive binary systems, which aims to constrain the internal structure of the stars. We focus on the binary CPD-41$^\circ$\,7742 and briefly revisit the case of HD\,152218.}{Independent studies of CPD-41$^\circ$\,7742 in the past showed large discrepancies in the longitude of periastron of the orbit, hinting at the presence of apsidal motion. We here perform a consistent analysis of all observational data, explicitly accounting for the rate of change of the longitude of periastron.}{We make use of the extensive set of spectroscopic and photometric observations of CPD-41$^\circ$\,7742 to infer values for the fundamental parameters of the stars and of the binary. Applying a disentangling method to the spectra allows us to simultaneously derive the radial velocities (RVs) at the times of observation and reconstruct the individual spectra of the stars.  The spectra are analysed by means of the \texttt{CMFGEN} model atmosphere code to determine the stellar properties. We determine the apsidal motion rate in two ways: First, we complement our RVs with those reported in the literature, and, second, we use the phase shifts between the primary and secondary eclipses. The light curves are further analysed by means of the \texttt{Nightfall} code to constrain the orbital inclination and, thereby, the stellar masses. Stellar structure and evolution models are then constructed with the \texttt{Cl\'es} code for the two stars with the constraints provided by the observations. Different prescriptions for the mixing inside the stars are adopted in the models. Newly available photometric data of HD\,152218 are analysed, and stellar structure and evolution models are built for the system as for CPD-41$^\circ$\,7742.}{The binary system CPD-41$^\circ$\,7742, made of an O9.5\,V primary ($M_\text{P}=17.8 \pm 0.5\,M_\odot$, $R_\text{P}=7.57 \pm 0.09\,R_\odot$, $T_\text{eff,P}=31\,800 \pm 1000$\,K, $L_\text{bol,P}=5.28^{+0.67}_{-0.68}\times 10^4\,L_\odot$) and a B1--2\,V secondary ($M_\text{S}=10.0 \pm 0.3\,M_\odot$, $R_\text{S}=4.29^{+0.04}_{-0.06}\,R_\odot$, $T_\text{eff,S}=24\,098 \pm 1000$\,K, $L_\text{bol,S}=5.58^{+0.93}_{-0.94} \times 10^3 \,L_\odot$), displays apsidal motion at a rate of $15\fdg38^{+0.42}_{-0.51}$\,yr$^{-1}$. Initial masses of $18.0\pm0.5\,M_\odot$ and $9.9\pm0.3\,M_\odot$ are deduced for the primary and secondary stars, respectively, and the binary's age is estimated to be $6.8\pm1.4$\,Myr. Regarding HD\,152218, initial masses of $20.6\pm 1.5$ and $15.5\pm 1.1$\,$M_\odot$ are deduced for the primary and secondary stars, respectively, and the binary's age of $5.2\pm0.8$\,Myr is inferred. }{Our analysis of the observational data of CPD-41$^\circ$\,7742 that explicitly accounts for the apsidal motion allows us to explain the discrepancy in periastron longitudes pointed out in past studies of this binary system. The age estimates are in good agreement with estimates obtained for other massive binaries in NGC\,6231. This study confirms the need for enhanced mixing in the stellar evolution models of the most massive stars to reproduce the observational stellar properties; this points towards larger convective cores than usually considered.  } 

\maketitle

\section{Introduction}
The high incidence of O-type binary systems in the young open cluster NGC\,6231 \citep[][and references therein]{San08} makes this cluster, located at the core of the Sco OB1 association, an interesting target for the study of massive star structure and evolution. Stars more massive than 8\,$M_\odot$ appear to be more concentrated in the cluster core \citep{Kuh17}. The ages of low-mass pre-main-sequence stars in NGC\,6231 range between 1 and 7\,Myr, with a small peak near 3\,Myr \citep{San07,Sung13}. More recently, the ages of three massive binary systems belonging to the cluster have been assessed: \citet{rauw16} derived an age estimate of $5.8\pm 0.6$\,Myr for HD\,152218, \citet{rosu20b} derived an age estimate of $5.25\pm 0.14$\,Myr for HD\,152248, and \citet{rosu22} derived an age estimate of $9.5\pm 0.6$\,Myr for HD\,152219, though the authors indicated that this latter age was probably overestimated. The common feature between these three studies is that the age is determined based on the analysis of the apsidal motion of these binaries. Indeed, the apsidal motion rate in an eccentric binary depends upon the internal structure constant of the stars, that is to say, the density stratification between the stellar core and the external layers, and hence, the evolutionary stage (i.e. the age) of the stars composing the system. This paper follows the path of these three studies and concentrates on a fourth massive binary system belonging to NGC\,6231, namely CPD-41$^\circ$\,7742. We also briefly revisit the HD\,152218 system in light of the newly available photometric data and stellar evolution codes that have been improved since our first analysis of this system \citep{rauw16}. 

In a broader context, the measurement of the apsidal motion rate in eccentric eclipsing binaries offers a unique opportunity to gain direct insight into the internal structure of the stars. Complemented by the fundamental stellar properties obtained from the light curves and from spectroscopy, the apsidal motion rate yields additional constraints that can be used to perform critical tests of stellar structure and evolution models \citep[e.g.][and references therein]{claret19, claret21}. There has been renewed interest in this field recently thanks to the Transiting Exoplanet Survey Satellite (TESS) mission \citep{ricker15} which provides densely covered light curves of eclipsing binary systems at very high precision \citep{baroch21}.

\begin{table*}[]
\caption{Physical and orbital parameters of CPD-41$^\circ$\,7742 from the literature.}
\centering
\begin{tabular}{l l l l l l l}
\hline\hline
\vspace{-3mm}\\
Parameter & \multicolumn{6}{c}{Value} \\
Reference& H74& LM83 & GM01 & S03 \& S05 & B05 & T10 \\
\hline
\vspace{-3mm}\\
$P_\text{orb}$ (d) & 2.446(5) & 2.430255(10) & 2.453087(12) & 2.44070(43) & 2.440656 & ... \\
$e$ & 0.07(4) & 0.08(3) & 0.205(53) & 0.027(6) & 0 & ... \\
$K_\text{P}$ (km\,s$^{-1}$) & 162.5(8.2) & 160(5) & 171(11) & 167.1(9) & 167.1(2.1) & ... \\
$K_\text{S}$ (km\,s$^{-1}$) & ... & ... & 364(59) & 301.3(1.8) & 299.9(3.8) & ... \\
$i$ & ... & ... & ... & 77\fdg35(80) & $81\fdg4(1)$ & ... \\
$M_\text{P}$ ($M_\odot$) & ... & ... & ... & 17.97(45) & 16.83(48) & 17.21(46) \\
$M_\text{S}$ ($M_\odot$) & ... & ... & ... & 9.96(22) & 9.38(27) & 9.59(27) \\
$R_\text{P}$ ($R_\odot$) & ... & ... & ... & 7.45(45) & 7.45(7) & 7.507(81) \\
$R_\text{S}$ ($R_\odot$) & ... & ... & ... & 5.39(43) & 4.18(4) & 4.217(89) \\
$T_\text{eff,P}$ (K) & ... & ... & ... & 34\,000 & 33\,200 &  33\,200(900) \\
$T_\text{eff,S}$ (K) & ... & ... & ... & 26\,260(420) & 26\,330(30) &  26\,330(900) \\
$\log g_\text{P}$ (cgs) & ... & ... & ... & 3.93(48) & 3.90 & 3.923(8) \\
$\log g_\text{S}$ (cgs) & ... & ... & ... & 3.96(64) & 4.16 &4.170(13) \\
$\log (L_\text{bol, P}/L_\odot)$ & ... & ... & ... & 4.82(7) & 4.785(8) & 4.789(48) \\
$\log (L_\text{bol, S}/L_\odot)$ & ... & ... & ... & 4.09(10) & 3.899(9) &3.885(62) \\
\vspace{-3mm}\\
\hline
\end{tabular}
\tablefoot{The references are the following: H74 \citep{Hil74}, LM83 \citep{Lev83}, GM01 \citep{Gar01}, S03 \citep{sana03}, S05 \citep{sana05}, B05 \citep{bouzid05}, and T10 \citep{torres10}. 
The parameters are the following: $P_\text{orb}$, the orbital period of the system determined from a best-fit phase-folding of the RV or photometric data not accounting for apsidal motion; $e$, the eccentricity of the orbit; $K_\text{P}$ (respectively $K_\text{S}$), the amplitude of the RV curve of the primary (respectively secondary) star; $i$, the orbital inclination; $M_\text{P}$ (respectively $M_\text{S}$), the mass; $R_\text{P}$ (respectively  $R_\text{S}$), the radius,  $T_\text{eff,P}$ (respectively $T_\text{eff,S}$), the effective temperature; $\log g_\text{P}$ (respectively $\log g_\text{S}$); the surface gravity; and $\log (L_\text{bol, P}/L_\odot)$ (respectively $\log (L_\text{bol, S}/L_\odot)$), the luminosity of the primary (respectively secondary) star. Errors are indicated in parentheses in terms of the last digits and represent $1\sigma$. }  
\label{table:intro}
\end{table*}

The massive binary CPD-41$^\circ$\,7742 -- also known as V\,1034 Sco -- is the second known eclipsing early-type double-lined spectroscopic (SB2) binary system of NGC\,6231, discovered after HD\,152248. \citet{struve44} was the first to suspect the binarity of CPD-41$^\circ$\,7742. With more data, \citet{Hil74} derived the first orbital solution for the system (see Table\,\ref{table:intro}) and reported a photometric variability of $\Delta V = 0.45$\,mag.  The primary star was assigned the spectral type O9\,IV by \citet{levato80}. \citet{Lev83} derived a new orbital solution (see Table\,\ref{table:intro}). Whilst \citet{Per90} published additional radial velocity (RV) data points, they did not present any new orbital solution for the system. \citet{Gar01} reported the first evidence of the presence of the secondary star in three out of their eight new medium-resolution spectra. They derived the first SB2 solution of the system (see Table\,\ref{table:intro}). However, \citet{sana03} reported errors in the Julian dates of some of the observations of \citet{Gar01}, and redetermined the orbital period and the eccentricity of the system based on 34 new data points that complemented the previous literature data available at that time (see Table\,\ref{table:intro}). They determined an O9\,III + B1\,III spectral classification for the system, even though they claimed that the stars would be main-sequence stars rather than giant stars considering the physical configuration of the system. \citet{sana05} reassessed the spectral classification considering optical photometric observations of the system and inferred O9\,V + B1-1.5\,V spectral types. They also derived the inclination of the system, as well as the masses and radii of the stars (see Table\,\ref{table:intro}). Based on multi-band photometry, \citet{bouzid05} derived orbital and physical parameters for the system (see Table\,\ref{table:intro}). They assumed a circular orbit and interpreted the small eccentricity found in spectroscopy as a possible result of circumstellar matter or other radiating material not connected to the orbital motion of the binary system. \citet{torres10} determined an O9\,V+B1.5\,V spectral classification for the system and inferred absolute parameters for the stars (see Table\,\ref{table:intro}). 
CPD-41$^\circ$\,7742 is a well-detached system, and it is unlikely that this system could have undergone any mass-exchange episode in its past \citep{sana05}.

In this article, we perform a detailed study of the apsidal motion of CPD-41$^\circ$\,7742. The set of observational data we use is introduced in Sect.\,\ref{sect:observations}. In Sect.\,\ref{sect:specanalysis} we perform the spectral disentangling, reassess the spectral classification of the stars and analyse the reconstructed spectra by means of the {\tt CMFGEN} model atmosphere code \citep{Hillier}. The RVs deduced from the spectral disentangling are combined with data from the literature in Sect.\,\ref{sect:omegadot} to establish values for the orbital period and mass ratio, notably. Photometric data of CPD-41$^\circ$\,7742 are analysed in Sect.\,\ref{sect:photom} by means of the {\tt Nightfall} binary star code. In Sect.\,\ref{sect:times_minima}, the phase shifts between the primary and secondary eclipses are used to infer the rate of apsidal motion and the orbital eccentricity of the system. The orbital and physical parameters of CPD-41$^\circ$\,7742 are summarised in Sect.\,\ref{sect:summary}. In Sect.\,\ref{sect:cles}, theoretical stellar structure and evolution models are built with the \texttt{Cl\'es} code and confronted to the observational parameters of the binary system. The massive binary system HD\,152218 -- also known as V\,1007\, Sco -- is reconsidered in Sect.\,\ref{sect:HD152218}. We provide our conclusions in Sect.\,\ref{sect:conclusion}.

\section{Observational data \label{sect:observations}}
\subsection{Spectroscopy}
To investigate the optical spectrum of CPD-41$^\circ$\,7742, we extracted 166 medium-high-resolution \'echelle spectra from the ESO science archive. Those spectra were collected between May 1999 and September 2017 using different instruments. Twenty-six of these were obtained with the Fiber-fed Extended Range Optical Spectrograph (FEROS) mounted on the European Southern Observatory (ESO) 1.5\,m telescope in La Silla, Chile \citep{Kaufer}, between May 1999 and April 2002 \citep{sana03}. We refer to \citet{sana03} for information about the instrumentation. 
The FEROS data were reduced using the FEROS pipeline of the {\tt MIDAS} software. Forty-one spectra were obtained with the High Accuracy Radial velocity Planet Searcher (HARPS) spectrograph attached to the Cassegrain focus of the ESO 3.6\,m telescope in La Silla during five consecutive nights in April 2009. The HARPS instrument has a spectral resolving power of 115\,000 \citep{mayor03}. Its detector consists of a mosaic of two CCD detectors that have altogether $4096 \times 4096$ pixels of $15\times 15$\,$\mu$m. The wavelength domain, covered by 72 orders, ranges from 3780 to 6910\,$\AA$, with a small gap between 5300 and 5330\,$\AA$. Exposure times of the observations were 300 seconds. The remaining 99 spectra were obtained with the GIRAFFE spectrograph mounted on the ESO Very Large Telescope (VLT) at Cerro Paranal \citep{pasquini02}, between April and September 2017. The GIRAFFE instrument has a spectral resolving power of 6300 and its EEV CCD detector has $2048 \times 4096$ pixels of $15\times 15$\,$\mu$m. The wavelength domain ranges from 3950 to 4572\,$\AA$. Exposure times of the observations range between 220 and 595 seconds. Both HARPS and GIRAFFE spectra provided in the archive were already reduced using the dedicated pipelines. Residual cosmic rays and telluric absorption lines were removed using {\tt MIDAS} and the {\tt telluric} tool within {\tt IRAF}, respectively.  We normalised the spectra with {\tt MIDAS} by fitting low-order polynomials to the continuum. The journal of the spectroscopic observations is presented in Appendix\,\ref{appendix:spectrotable}, Table\,\ref{Table:spectro+RV}.

\subsection{Photometry\label{subsect:obs_photo}}
Several sets of photometric data are available for CPD-41$^\circ$\,7742.
Between 22 March and 19 April 1997, the binary system was observed with the 0.6\,m Bochum telescope at La Silla observatory. Data were collected with two narrow-band filters, one centred on the He\,{\sc ii} $\lambda$\,4686 line, the other one centred at 6051\,$\AA$ \citep{Royer}. In the following, we refer to these data as Bochum 4686 and Bochum 6051, respectively. These data were previously used by \citet{sana05}, and we refer to this paper for details about the observations and the instrument characteristics.  

The second set of photometric observations consists of $uvby$ Str\"omgren data obtained at ESO/La Silla with the Danish 1.54\,m telescope equipped with either a LORAL $2048\times 2048$ CCD chip (observations between June and July 2000) or an EEV/MAT $2048\times 4096$ CCD chip (observations between March and July 2001) in the framework of the Long-term Photometry of Variables project \citep{sterken83, sterken94}. Data reduction was carried out by \citet{bouzid05} and we refer to this paper for more detailed information.  

The system was also recently observed by TESS during sectors 12 (i.e. between 21 May and 19 June 2019, hereafter TESS-12) and 39 (i.e. between 27 May and 24 June 2021, hereafter TESS-39). The TESS detector bandpass spans from 600 to 1000\,nm and is centred on the Cousins $I$-band with central wavelength 786.5\,nm. The cadence was 30\,min for TESS-12 and 10\,min for TESS-39. We built light curves for this system using aperture photometry performed with the Python package \texttt{Lightkurve}\footnote{https://docs.lightkurve.org/}. The source extraction was done in the single central pixel of a $50\times50$ pixel image (cut-out) to avoid contamination from neighbouring sources as much as possible. As a background mask, we used pixels with fluxes below the median flux. Several corrections for background contamination were tried: a simple median of the background pixels as well as a principal component analysis with two (pca-2) or five (pca-5) components. Only the first one could be done for TESS-12 while all three methodologies led to similar results for TESS-39 and we adopted the pca-5 result in this case. All data points with errors larger than the mean of the errors plus three times their 1$\sigma$ dispersion were discarded. The TESS fluxes were converted into magnitudes and the mean magnitude in each sector was subtracted. The time intervals that showed a long-term (instrumental) trend were discarded: For TESS-12, we removed the data taken during the first half of the sector (i.e. before the satellite's perigee gap); For TESS-39, we skipped data taken within about two days of the beginning or end of the sector and around the perigee gap. 

\section{Spectral analysis\label{sect:specanalysis}}
\subsection{Spectral disentangling\label{subsect:spectraldisentangling}}
We applied our disentangling code based on the method described by \citet{GL} on all data to derive the individual spectra of the binary components as well as their RVs at the times of the observations. The method is described in details in \citet{Rosu, rosu22} to which we refer here for more information about the method and its limitations. 

For the synthetic {\tt TLUSTY} spectra, used as cross-correlation templates in the determination of the RVs, we assumed $T_\text{eff} = 30\,000$\,K, $\log g = 3.75 $, and $v \sin i_\text{rot} =  160$\,km\,s$^{-1}$ for the primary star and $T_\text{eff} = 21\,000$\,K, $\log g = 4.00$, and $v \sin i_\text{rot} = 160$\,km\,s$^{-1}$ for the secondary star.

We adopted the following separate wavelength domains: A0[3810:3920]\,\AA, A1[3990:4400]\,\AA, A2[4300:4565]\,\AA, A3[4570:5040]\,\AA, A4[5380:5860]\,\AA, A5[5832:5885]\,\AA, A6[6450:6720]\,\AA,~and A7[7025:7095]\,\AA. 
We first processed the wavelength domains (A0, A1, A2, A3, A6, and A7) for which the code was able to reproduce the individual spectra and simultaneously estimate the RVs of the stars. We note, however, that only FEROS spectra cover the A7 wavelength domain, while GIRAFFE spectra only cover the A1 and A2 domains. The RVs from the individual wavelength domains were then averaged using weights corresponding to the number of strong lines present in these domains (three lines for A0, six for A1, three for A2, four for A3, two for A6, and one for A7). The resulting RVs of both stars are reported in Table\,\ref{Table:spectro+RV} together with their $1\sigma$ errors. We finally performed the disentangling on the remaining two domains (A4 and A5) with the RVs fixed to these weighted averages, and, for the A4 domain only, due to the presence of interstellar medium lines, using a version of the code designed to deal with a third spectral component \citep{Mahy12}. 

\subsection{Spectral classification and absolute magnitudes\label{subsect:spectralclass}}
The reconstructed spectra of the binary components of CPD-41$^\circ$\,7742 allowed us to reassess the spectral classification of the stars. 

For the primary star, we used Conti's criterion \citep{Conti71} complemented by \citet{Mathys88} to determine the spectral type of the star. We found that $\log W' = \log[\text{EW(He\,{\sc i} $\lambda$\,4471)/EW(He\,{\sc ii} $\lambda$\,4542)}]$ amounts to $0.58\pm 0.01$, which corresponds to spectral type O9.5. Hereafter, errors are given as $\pm1\sigma$. We complemented this criterion with Sota's criteria \citep{Sota11,Sota} based on the ratios between the intensities of several lines: The ratio He\,{\sc ii} $\lambda$\,4542/He\,{\sc i} $\lambda$\,4388 amounts to $0.66\pm0.01$, the ratio He\,{\sc ii} $\lambda$\,4200/He\,{\sc i} $\lambda$\,4144 amounts to $0.75\pm0.01$, and the ratio Si\,{\sc iii} $\lambda$\,4552/He\,{\sc ii} $\lambda$\,4542 amounts to $0.44\pm0.01$. All three ratios being lower than unity, the spectral type O9.7 is clearly excluded and the spectral type O9.5 is confirmed. 
To assess the luminosity class of the primary star, we used Conti's criterion \citep{Conti71} and found that $\log W' = \log[\text{EW(Si\,{\sc iv} $\lambda$\,4089)/EW(He\,{\sc i} $\lambda$\,4143)}]$ amounts to $0.08\pm 0.01$, which corresponds to the main-sequence luminosity class V. This is confirmed by the appearance of the spectral lines Si\,{\sc iv} $\lambda$\,4116 and He\,{\sc i} $\lambda$\,4121 of the same strength and weaker than H$\delta$ in line with expectations for luminosity class V (see the spectral atlas of \citet{Gray} for spectral type O9).  

For the secondary star, we used the spectral atlas of \citet{Gray}. As He\,{\sc ii} lines are absent in the spectrum, spectral type O is excluded. Qualitatively, the ratio between the strengths of the He\,{\sc i} $\lambda$\,4471 and Mg\,{\sc ii} $\lambda$\,4481 lines suggests a spectral type between B1\,V and B3\,V. This statement is confirmed by the fact that the Si\,{\sc ii} $\lambda\lambda$ 4128-32 lines are only marginally present but should be clearly visible for spectral types later than B2.5. In addition, the Si\,{\sc iv} $\lambda$\,4089 line is clearly visible but weaker than the Si\,{\sc iii} $\lambda$\,4552 line, suggesting a spectral type B1, with an uncertainty of one spectral type. However, using the spectral atlas of \citet{WP90}, a B2 V class is favoured as the strengths of the He\,{\sc i} $\lambda$\,4387 and 4471 lines are nearly identical, the strength of the O\,{\sc ii} $\lambda$\,4416 line is much weaker than the strength of the He\,{\sc i} $\lambda$\,4387 line, and the strength of the He\,{\sc i} $\lambda$\,4713 line exceeds those of the O\,{\sc ii} $\lambda\lambda$\,4639-42 and N\,{\sc ii} $\lambda$\,4630 and C\,{\sc iii} $\lambda$\,4647 lines. Regarding the luminosity class, the ratio of the strengths of the Si\,{\sc iii} $\lambda$\,4552 and He\,{\sc i} $\lambda$\,4387 lines suggests a main-sequence star (class V). Furthermore, the O\,{\sc ii} $\lambda$\,4348 line is barely visible, the O\,{\sc ii} $\lambda$\,4416 line is weak and the Balmer lines are wide, which confirms that the secondary star is of luminosity class V. In conclusion, the secondary star is a B1--2\,V star. 

The brightness ratio in the visible domain was estimated based on the ratio between the equivalent widths (EWs) of the spectral lines of the  secondary star and {\tt TLUSTY} spectra of similar effective temperatures. For this purpose, we used the H$\beta$, He\,{\sc i} $\lambda\lambda$\,4026, 4144, 4471, 4921, 5016, and 5876 lines. 
The ratio $\text{EW}_{\text{\tt TLUSTY}}/\text{EW}_{\text{sec}} = (l_\text{P}+l_\text{S})/l_\text{S}$ is equal to $6.73\pm1.02$, $6.62\pm0.94$, $6.42\pm0.89$, $6.24\pm0.97$, and $5.97\pm0.87$ for $T_\text{eff}$ of 20\,000, 21\,000, 22\,000, 23\,000, and 24\,000\,K, respectively. As these values are very similar, we took the value for the {\tt TLUSTY} spectrum of 22\,000\,K, which corresponds to the temperature of a B1.5\,V star as inferred by \citet{humphreys}, and derived a value for $l_\text{S}/l_\text{P}$ of $0.18\pm0.03$. \\

The {\it Gaia} early data release 3 \citep[EDR3,][]{EDR3} quotes a parallax of $\varpi = 0.6452 \pm 0.0231$\,mas, corresponding to a distance of $1489^{+64}_{-61}$\,pc \citep{bailer21}. From this distance, we derived a distance modulus of $10.87\pm 0.09$. \citet{Zacharias13} reported mean $V$ and $B$ magnitudes of $8.80$ and $9.02$, respectively. For these two values, we estimated errors of 0.01. Adopting a value of $-0.26\pm0.01$ for the intrinsic colour index $(B-V)_0$ of an O9.5\,V star \citep{MP} and assuming the reddening factor in the $V$-band $R_V$ equal to $3.3\pm0.1$ for NGC\,6231 \citep{Sung98}, we obtained an absolute magnitude in the $V$-band of the binary system $M_V=-3.65\pm0.12$. The brightness ratio then yields individual absolute magnitudes $M_{V,\text{P}} =-3.47\pm0.12$ and $M_{V,\text{S}} =-1.63\pm0.19$ for the primary and secondary stars, respectively. 
 
Comparing the magnitude obtained for the primary star to those reported by \citet{MP} shows that the primary is slightly less luminous than a `typical' O9.5\,V star. Likewise, comparing the magnitude obtained for the secondary star to those reported by \citet{humphreys} shows that the secondary is fainter than expected for B1--2\,V type stars. These comparisons clearly rule out luminosity class II-III for both stars.

\subsection{Projected rotational velocities\label{subsect:vsini}}
\begin{figure*}[h!]
\centering
\includegraphics[width=0.49\linewidth, trim=2cm 0cm 3cm 6.5cm, clip=true]{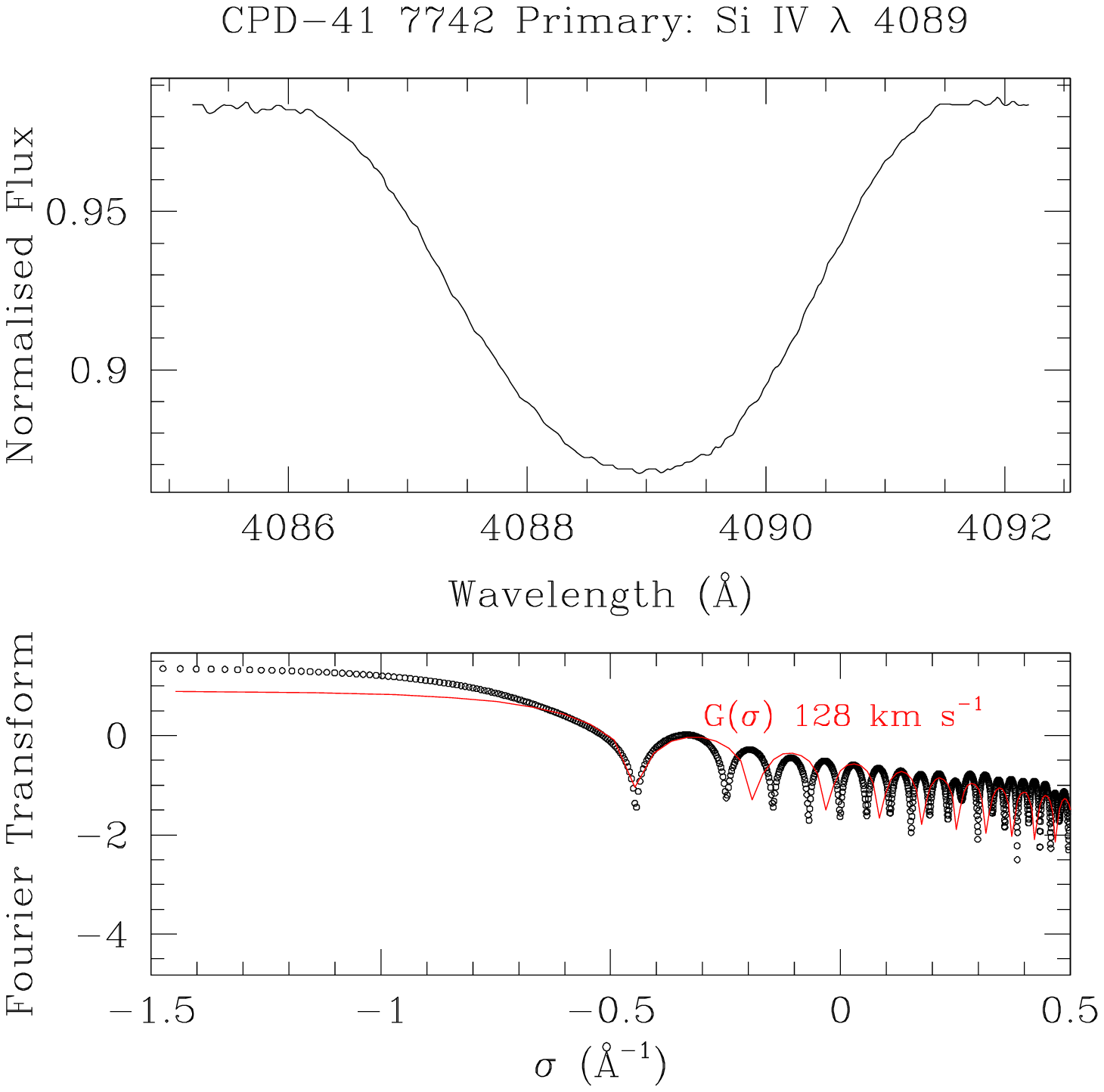}
\includegraphics[width=0.49\linewidth, trim=2cm 0cm 3cm 6.5cm, clip=true]{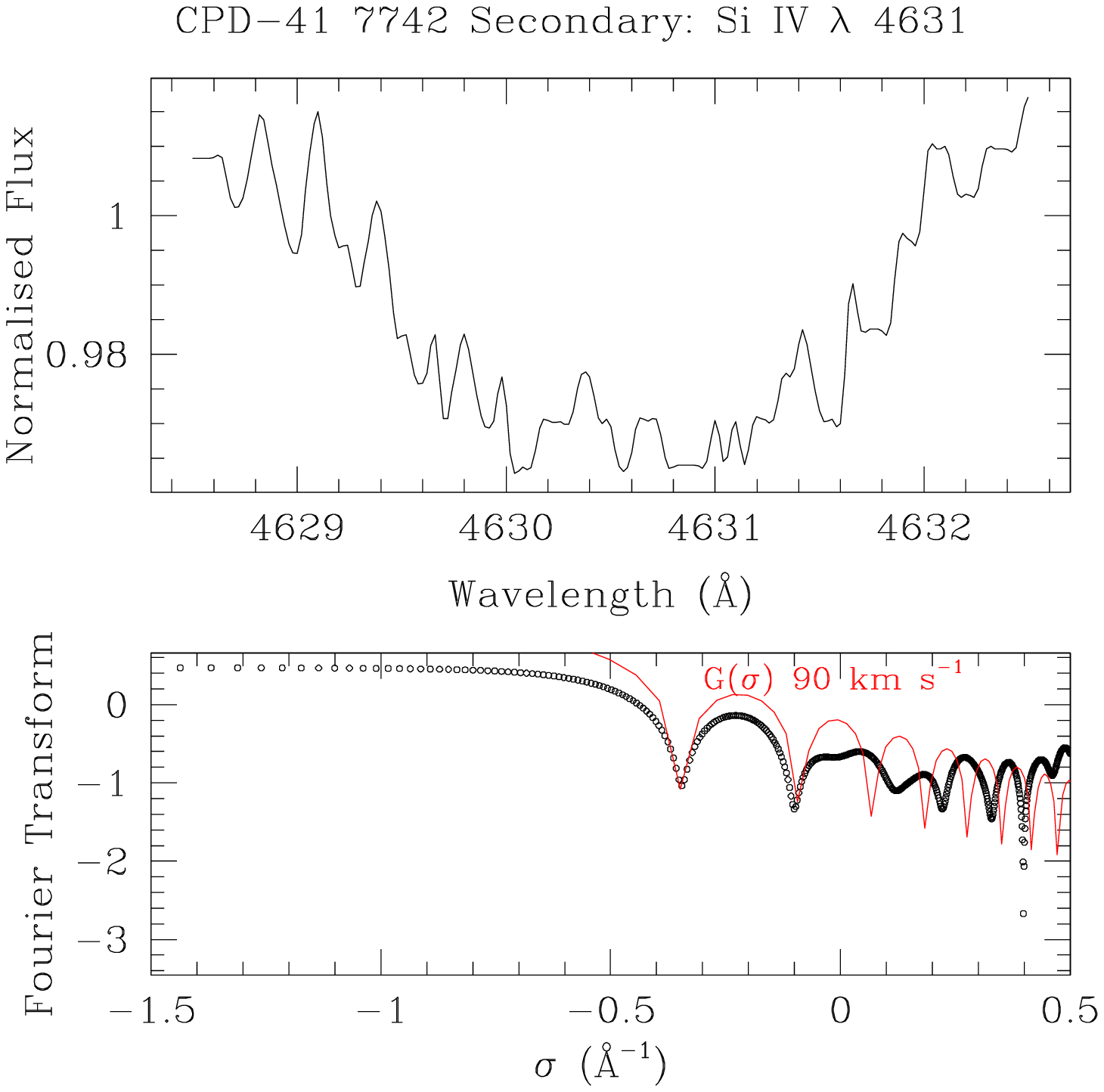}
\caption{Fourier transforms of primary and secondary lines. \textit{Top row:} Line profiles of the separated spectra of CPD-41$^\circ$\,7742 obtained after application of the brightness ratio for the primary (Si\,{\sc iv} $\lambda$\,4089 line, \textit{left panel}) and secondary (Si\,{\sc iv} $\lambda$\,4631 line, \textit{right panel}) stars. \textit{Bottom row:} Fourier transform of those lines (in black) and best-match rotational profile (in red) for the primary (\textit{left panel}) and secondary (\textit{right panel}) stars. \label{fig:vsini}}
\end{figure*}

The projected rotational velocities of both stars were derived using the Fourier transform method \citep{Simon-Diaz,Gray08}. We proceeded as in \citet{rosu22}. 
The results are presented in Table\,\ref{vsiniTable}, and the Fourier transforms of the Si\,\textsc{iv} $\lambda$\,4089 line for the primary star and of the Si\,\textsc{iv} $\lambda$\,4631 line for the secondary star are illustrated in Fig.\,\ref{fig:vsini}. 
The results presented in Table\,\ref{vsiniTable} show that the mean $v \sin i_\text{rot}$ computed on metallic lines alone or on all the lines agree very well. We adopt a mean $v \sin i_\text{rot}$ of $140\pm7$ km\,s$^{-1}$ for the primary star and of $89\pm7$ km\,s$^{-1}$ for the secondary star. The value of $v\sin i_\text{rot}$ for the primary star agrees with the value obtained by \citet{Lev83}.

\begin{table}[h!]
\caption{Best-fit projected rotational velocities as derived from the disentangled spectra of CPD-41$^\circ$\,7742.}
\centering
\begin{tabular}{l l l}
\hline\hline
\vspace{-3mm}\\
Line & \multicolumn{2}{c}{$v\sin i_{\text{rot}}$ (km\,s$^{-1}$)} \\
& Primary & Secondary \\
\hline
\vspace{-3mm}\\
Si\,{\sc iv} $\lambda$\,4089 & 128 & ... \\ 
Si\,{\sc iv} $\lambda$\,4212 & 132 & ... \\ 
O\,{\sc ii} $\lambda$\,4254 & 133 & ... \\ 
C\,{\sc ii} $\lambda$\,4267 & 134 & ... \\ 
Mg\,{\sc ii} $\lambda$\,4481 & ... & 89 \\ 
Si\,{\sc iv} $\lambda$\,4631 & ... & 90 \\ 
N\,{\sc iii} $\lambda$\,4641 & ... & 98 \\ 
O\,{\sc iii} $\lambda$\,5592 & 146 & ... \\ 
He\,{\sc i} $\lambda$\,4026 & 140 & ... \\ 
He\,{\sc i} $\lambda$\,4143 & 135 & ... \\ 
He\,{\sc i} $\lambda$\,4713 & 146 & 88 \\ 
He\,{\sc i} $\lambda$\,4922 & 144 & 88 \\ 
He\,{\sc i} $\lambda$\,5016 & 146 & 84 \\
He\,{\sc i} $\lambda$\,5876 & 146 & 86 \\
He\,{\sc i} $\lambda$\,6678 & 146 & 77 \\
He\,{\sc i} $\lambda$\,7065 & 144 & 100 \\
\hline
\vspace{-3mm}\\
Mean (metallic lines) & $135\pm7$ &$92\pm5$ \\ 
Mean (He\,{\sc i} lines) & $143\pm4$ &$87\pm7$ \\ 
Mean (all lines) & $140\pm7$ & $89\pm7$ \\
\hline
\vspace{-3mm}\\
\citet{Lev83} & $130$ & ...\\
\hline
\end{tabular}
\tablefoot{The values quoted by \citet{Lev83} were obtained by visual comparison with the standards given by \citet{Slettebak}.}
\label{vsiniTable}
\end{table} 

\subsection{Model atmosphere fitting\label{subsect:cmfgen}}
The reconstructed spectra of the binary components were analysed by means of the {\tt CMFGEN} model atmosphere code \citep{Hillier} to constrain the fundamental properties of the stars. We here refer the reader to \citet{rosu22} for information about the code and its limitations. 

The {\tt CMFGEN} spectra were first broadened by the projected rotational velocities determined in Sect.\,\ref{subsect:vsini}. We adjusted the stellar and wind parameters following the procedure outlined by \citet{Martins11}.  

\subsubsection{Primary star}
The macroturbulence velocity was adjusted on the wings of the O\,{\sc iii} $\lambda$\,5592 and Balmer lines, and we derived a value of $v_\text{macro}=100\pm 10$\,km\,s$^{-1}$. We adjusted the microturbulence velocity, that is to say, the microturbulence value at the level of the photosphere, $v_\text{micro}^\text{min}$, on the metallic lines and inferred a value of $14\pm3$\,km\,s$^{-1}$.

The effective temperature was determined by searching for the best overall fit of the He\,{\sc i} and He\,{\sc ii} lines. This was clearly a compromise as we could not find a solution that perfectly fits all helium lines simultaneously. For instance, we found that the strength of the He\,{\sc i} $\lambda$\,4471 line is underestimated while other He\,{\sc i} lines are well-adjusted. Given the luminosity class of the primary, it seems unlikely that this discrepancy reflects the dilution effect discussed by \citet{voels89} and \citet{herrero92}. We discarded some lines that are only weakly sensitive to temperature (He\,{\sc i} $\lambda$\,4713), or that are blended with neighbouring lines (He\,{\sc i} $\lambda$\,4026 is blended with the weak but non-zero He\,{\sc ii} $\lambda$\,4026 line and He\,{\sc i} $\lambda$\,4121 is blended with Si\,{\sc iv} $\lambda$\,4116). Likewise, the He\,{\sc ii} $\lambda$\,5412 line could be impacted by the stellar wind \citep{herrero92}. Hence, we adjusted the effective temperature on the He\,{\sc i} $\lambda$\,4922 line and obtained a value of $31\,800\pm 1000$\,K. With this effective temperature, the He\,{\sc i} $\lambda\lambda$\,4026, 4713, and 4922 lines as well as the He\,{\sc ii} $\lambda$\,4200 line are well adjusted. 

The surface gravity was obtained by adjusting the wings of the Balmer lines H$\beta$, H$\gamma$, H$\delta,$ H\,{\sc i} $\lambda\lambda$\,3890, and 3835. We derived $\log g_\text{spectro} = 3.76\pm0.10$.  

\begin{figure*}[htbp]
\centering
\includegraphics[clip=true,trim=0.5cm 4.5cm 1cm 1cm,width=\linewidth]{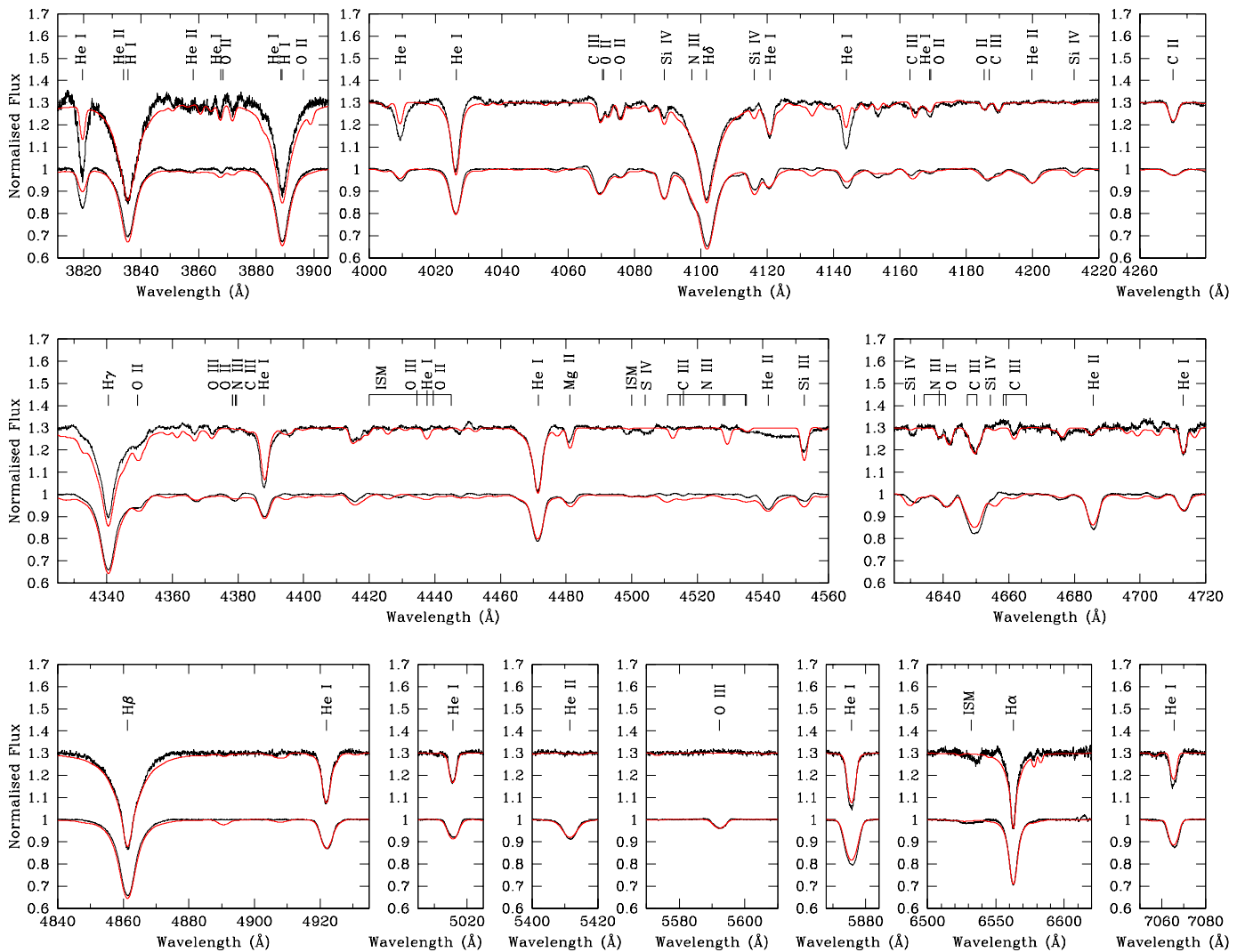}
\caption{Normalised disentangled spectra (in black) of the primary and secondary stars of CPD-41$^\circ$\,7742 (the spectrum of the secondary star is shifted by +0.3 in the $y$-axis for clarity) together with the respective best-fit {\tt CMFGEN} model atmosphere (in red). \label{Fig:CMFGEN}}
\end{figure*}

The surface chemical abundances of all elements, including helium, carbon, nitrogen, and oxygen, were set to solar as taken from \citet{Asplund}. The primary star of CPD-41$^\circ$\,7742 is a typical case where it is impossible to certify whether or not the chemical abundances of the elements differ from solar; the number of parameters we have to adjust exceeds the number of lines we can use for this purpose.
For instance, the oxygen abundance could be determined by adjusting the O\,{\sc iii} $\lambda$\,5592 line as this is the sole oxygen line free of blends. The best fit is obtained for a sub-solar abundance O/H of $(4.20 \pm0.20)\times10^{-4}$. The synthetic {\tt CMFGEN} spectrum displays a series of weak O\,{\sc iii} absorption lines ($\lambda\lambda$\,4368, 4396, 4448, 4454, and 4458) that are not present in the observed spectra (neither before nor after disentangling). These lines are not correctly reproduced by the model, as has been pointed out by several authors \citep[e.g.][]{Raucq,Rosu,rosu22}, and hence, are not considered in our analysis. The {\tt CMFGEN} spectrum also displays the O\,{\sc iii} $\lambda$\,5508 line, which is not present in the stellar spectrum whereas the O\,{\sc ii} $\lambda$\,4070 line is correctly reproduced. However, this line is blended with C\,{\sc iii} $\lambda$\,4070 and should therefore be considered with caution. A similar caveat applies to the O\,{\sc ii}\, $\lambda$\,4185 line that is heavily blended with the C\,{\sc iii} $\lambda$\,4187 line. \\
Similarly, the carbon abundance could in principle be adjusted on the C\,{\sc iii} $\lambda$\,4070 and C\,{\sc ii} $\lambda$\,4267 lines. 
However, the C\,{\sc iii} $\lambda$\,4070 line is blended with O\,{\sc ii} $\lambda$\,4070. We note that the C\,{\sc iii} $\lambda\lambda$\,4647-51 blend and the C\,{\sc iii} $\lambda$\,5696 line are known to be problematic because their formation processes are controlled by a number of far-UV lines \citep{Martins12}. As a result, the strength and nature of the C\,{\sc iii} $\lambda\lambda$\,4647-51 blend and of the C\,{\sc iii} $\lambda$\,5696 line critically depend upon subtle details of the stellar atmosphere model. \\
Finally, the nitrogen abundance is impossible to determine with accuracy as the N\,{\sc iii} $\lambda\lambda$\,4510-4540 blend is completely absent from the stellar spectrum, the $\lambda\lambda$\,4634-4640 complex is heavily blended with Si\,{\sc iv}, O\,{\sc ii,} and C\,{\sc iii} lines, and the N\,{\sc iii} $\lambda$\,4379 line is heavily blended with the C\,{\sc iii} $\lambda$\,4379 line.  

Regarding the wind parameters, the clumping parameters were fixed: The volume filling factor $f_1$ was set to 0.1, and the $f_2$ parameter controlling the onset of clumping was set to 100\,km\,s$^{-1}$. For the sake of completeness, we varied $f_1$ from 0.05 to 0.2 and we did not observe any significant difference as the resulting spectra perfectly overlapped. 
Likewise, the $\beta$ parameter of the velocity law was fixed to 1.1 as suggested by \citet{Muijres} for an O9.5\,V type star. Again for the sake of completeness, we tested a value of 1.0 for $\beta$ but did not observe any significant differences  as the resulting spectra perfectly overlapped.

In principle, the wind terminal velocity could be derived from the H$\alpha$ line. However, because of the degeneracy between the wind parameters, we decided to fix the value of $v_\infty$ to 2380\,km\,s$^{-1}$ as measured by \citet{sana05}. Regarding the mass-loss rate, the main diagnostic lines in the optical domain are H$\alpha$ and He\,{\sc ii} $\lambda$\,4686. We measured a value of $\dot{M} = (5.0\pm1.0)\times10^{-8}\,M_\odot\,\text{yr}^{-1}$ based on the H$\alpha$ line but could not reproduce the He\,{\sc ii} $\lambda$\,4686 line correctly. \\

The stellar and wind parameters of the best-fit {\tt CMFGEN} model atmosphere are summarised in Table\,\ref{Table:CMFGEN}. The normalised disentangled spectra of the primary of CPD-41$^\circ$\,7742 are illustrated in Fig.\,\ref{Fig:CMFGEN} along with the best-fit {\tt CMFGEN} adjustment.    \\

\begin{table}[h!]
\caption{Stellar and wind parameters of the best-fit {\tt CMFGEN} model atmosphere derived from the separated spectra of CPD-41$^\circ$\,7742.}
\centering
\begin{tabular}{l l l}
\hline\hline
\vspace{-3mm}\\
Parameter & \multicolumn{2}{c}{Value} \\
& Primary & Secondary \\
\hline
\vspace{-3mm}\\
$T_\text{eff}$ (K) & $31\,800\pm1000$ & $26\,000\pm1000$  \\ 
$\log g_\text{spectro}$ (cgs) & $3.76\pm0.10$ & $4.00\pm0.10$ \\
$v_\text{macro}~(\text{km\,s}^{-1})$ & $100\pm10$ & $50\pm10$ \\ 
$v_\text{micro}^\text{min}~(\text{km\,s}^{-1})$ & $14\pm3$ & $10\pm3$ \\[0.1cm]
$\dot{M}~(M_\odot\,\text{yr}^{-1})$  & $(5.0\pm1.0)\times10^{-8}$ &$(5.0\pm1.0)\times 10^{-10}$ \\
$\dot{M}_\text{uncl.}~(M_\odot\,\text{yr}^{-1})^a$  & $(1.6\pm 0.3)\times10^{-7}$ & $(1.6\pm 0.3)\times10^{-9}$ \\
$v_\infty~(\text{km\,s}^{-1})$ & 2380 (fixed)  & $2040$ \\
$f_1$ & 0.1 (fixed) & 0.1 (fixed)  \\ 
$f_2~(\text{km\,s}^{-1})$ & 100 (fixed) & 100 (fixed)\\
$\beta$ & 1.1 (fixed) & 1.1  (fixed) \\
\vspace*{-3mm}\\
\hline
\end{tabular}
\tablefoot{ $^a$$\dot{M}_\text{uncl.}=\dot{M}/\sqrt{f_1}$ is the unclumped mass-loss rate.}
\label{Table:CMFGEN}
\end{table} 

\subsubsection{Secondary star} 
We proceeded differently for the secondary star. As the O\,{\sc iii} $\lambda$\,5592 line, the main diagnostic line for the macroturbulence velocity, is not present in the spectrum of the secondary star, we could not adjust $v_\text{macro}$ on this line and had to rely on He\,{\sc i} and Balmer lines instead. Hence, we adjusted $v_\text{macro}$, the effective temperature, and the surface gravity simultaneously in an iterative manner in order to get the best fit of the spectrum. We derived a value of  $v_\text{macro}=50\pm 10$\,km\,s$^{-1}$. 

The effective temperature was determined based on the sole He\,{\sc i} lines, as the secondary spectrum lacks He\,{\sc ii} lines. As for the primary star, we adjusted the effective temperature mainly on the He\,{\sc i} $\lambda$\,4922 line, and found a value of $26\,000 \pm 1000$\,K. With this effective temperature, the He\,{\sc i} $\lambda\lambda$\,4471, 4713, and 5016 lines are also well adjusted. However, the He\,{\sc i} $\lambda$\,4026 line is slightly overestimated while the He\,{\sc i} $\lambda\lambda$\,4388, 5876, and 7065 lines are slightly underestimated. 

The surface gravity was obtained by adjusting the wings of the Balmer lines H$\beta$, H$\gamma$, H$\delta$, and H\,{\sc i} $\lambda\lambda$\,3835, 3890. We derived $\log g_\text{spectro} = 4.00\pm0.10$. With this value, the wings of H$\beta$, H$\delta$, and H\,{\sc i} $\lambda$\,3835 are well adjusted while the wings of the two other lines are overestimated. 

We adjusted the microturbulence velocity on the metallic lines and obtained $10\pm3$\,km\,s$^{-1}$ for $v_\text{micro}^\text{min}$.
The surface chemical abundances of all elements, including He, C, N, and O were set to solar according to \citet{Asplund} for the same reasons as for the primary star. 

Regarding the wind parameters, the clumping parameters $f_1$ and $f_2$ were fixed as for the primary star. We fixed the $\beta$ parameter of the velocity law to 1.1 as we have no indication about the value of this parameter and did not find any significant difference when varying the value of this parameter. 

We adjusted a mass-loss rate of $\dot{M} = (5.0\pm1.0)\times10^{-10}\,M_\odot/\text{yr}$ based on the H$\gamma$ and H$\alpha$ lines. The wind terminal velocity was then adjusted based on the strength of the H$\alpha$ line. We found a value of  $v_\infty = 2040$\,km\,s$^{-1}$.

The stellar and wind parameters of the best-fit {\tt CMFGEN} model atmosphere are summarised in Table\,\ref{Table:CMFGEN}. The normalised disentangled spectra of the secondary of CPD-41$^\circ$\,7742 are illustrated in Fig.\,\ref{Fig:CMFGEN} along with the best-fit {\tt CMFGEN} adjustment.

\section{Radial velocity analysis \label{sect:omegadot}}
The total dataset of RVs consists of our 166 primary and secondary RVs determined via disentangling (see Sect.\,\ref{subsect:spectraldisentangling}) complemented by 37 primary and nine secondary RVs coming from the literature. For the primary star, there are 16 RVs from \citet{Hil74}, four from \citet{Lev83}, three from \citet{Per90}, eight from \citet{Gar01}, and six from \citet[their CAT data]{sana03}. The secondary RVs come from \citet{Gar01}, three values, and from \citet{sana03}, six values. As reported by \citet{sana03}, some of the Julian dates quoted by \citet{Gar01} were wrong by one day. These were corrected and reported by \citet{sana03} and we therefore adopted these corrected dates in the present analysis. In total, we ended up with a series of 203 primary RV data points spanning about 49 years and 175 secondary RV data points spanning about 22 years. We adopted the following values for the uncertainties on the primary RVs: 10\,km\,s$^{-1}$ for the data of \citet{Hil74}, \citet{Lev83}, and \citet{Gar01}, 15\,km\,s$^{-1}$ for the data of \citet{Per90}, and  5\,km\,s$^{-1}$ for the CAT data of \citet{sana03}. For the RVs derived from the spectral disentangling, we adopted formal errors of 5\,km\,s$^{-1}$, as the small errors derived as part of the disentangling method would bias the adjustment given our high number of RVs compared to those coming from the literature. Since the secondary RVs are only available for the most recent data and are subject to larger errors than the primary RVs, we did not use them in the determination of the rate of apsidal motion. 

For each time of observation $t$, we adjusted the primary RV data as in \citet{rosu22}, explicitly accounting for the apsidal motion through the variation of the argument of periastron of the primary orbit. The systemic velocity $\gamma_\text{P}$ of each subset of our total dataset was adjusted so as to minimise the sum of the residuals, because RVs of different spectral lines potentially yield slightly different apparent systemic velocities.

\begin{figure*}[h!]
\centering
\hspace{-6.93cm}\includegraphics[clip=true,trim=0cm 2.3cm 0.8cm 7cm, width=0.615\linewidth]{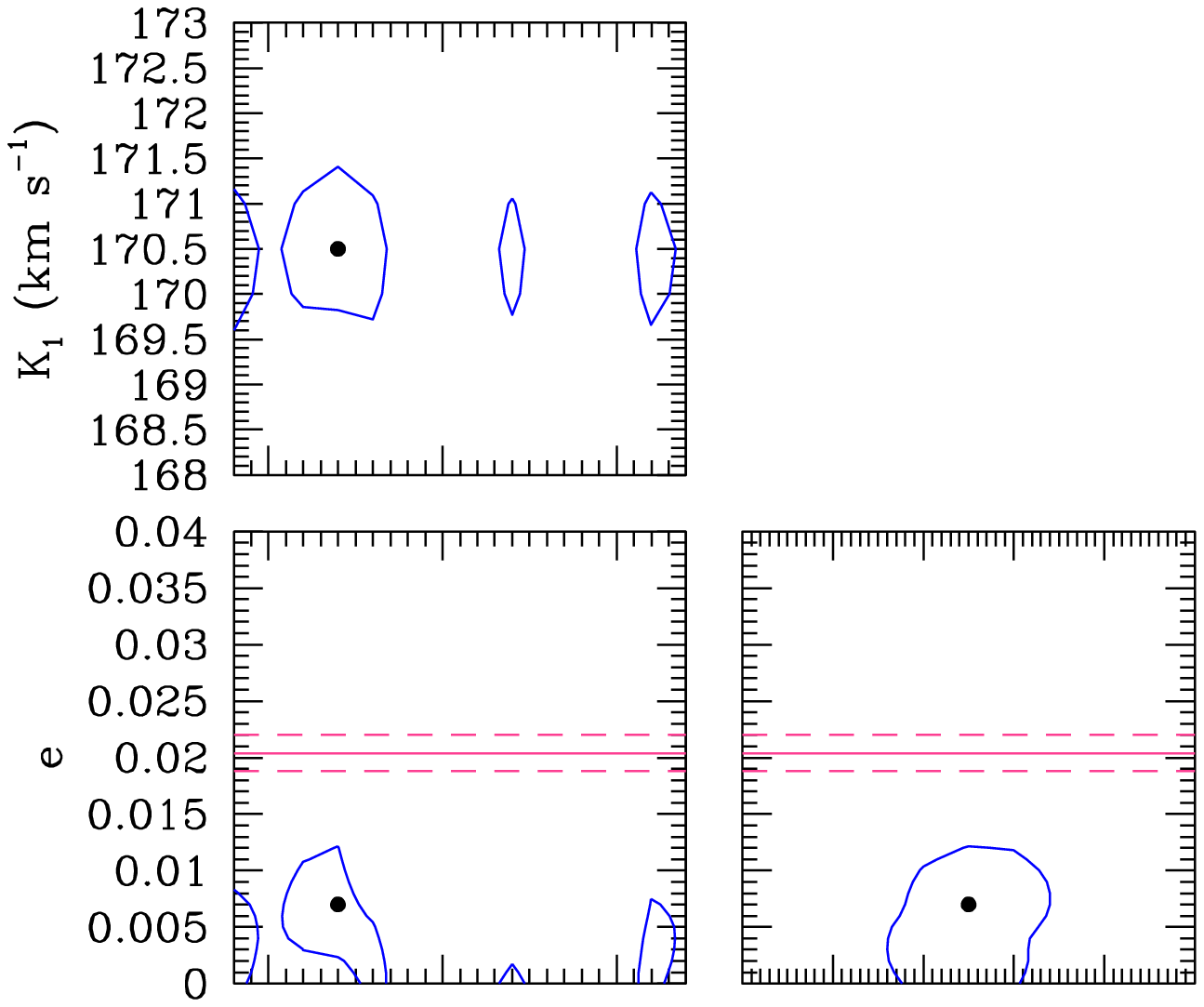}\\
\vspace{-0.7cm}
\includegraphics[clip=true,trim=0cm 1cm 0.8cm 1.4cm, width=0.615\linewidth]{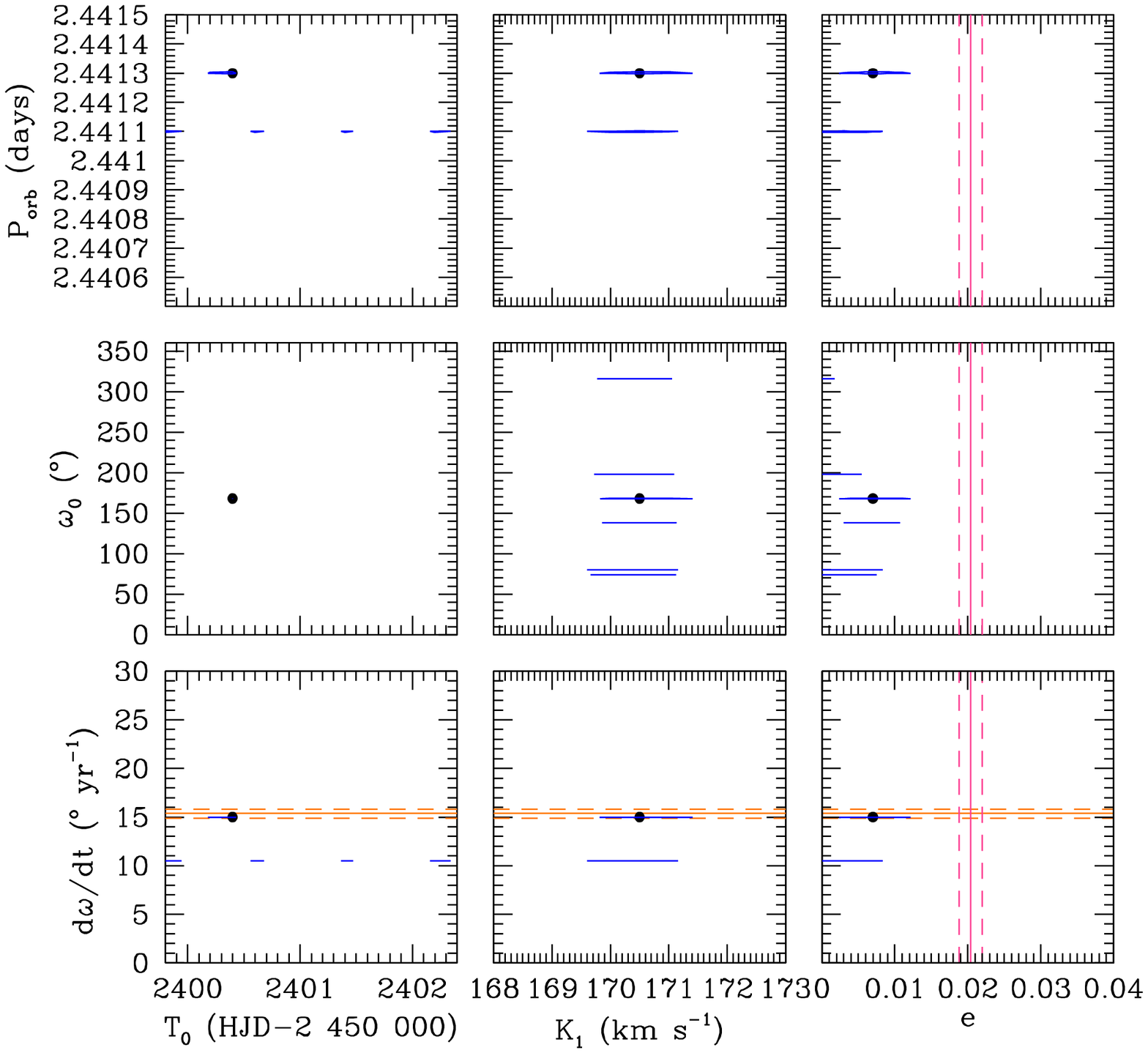}
\includegraphics[clip=true,trim=2.8cm 1cm 5.57cm 0cm, width=0.373\linewidth]{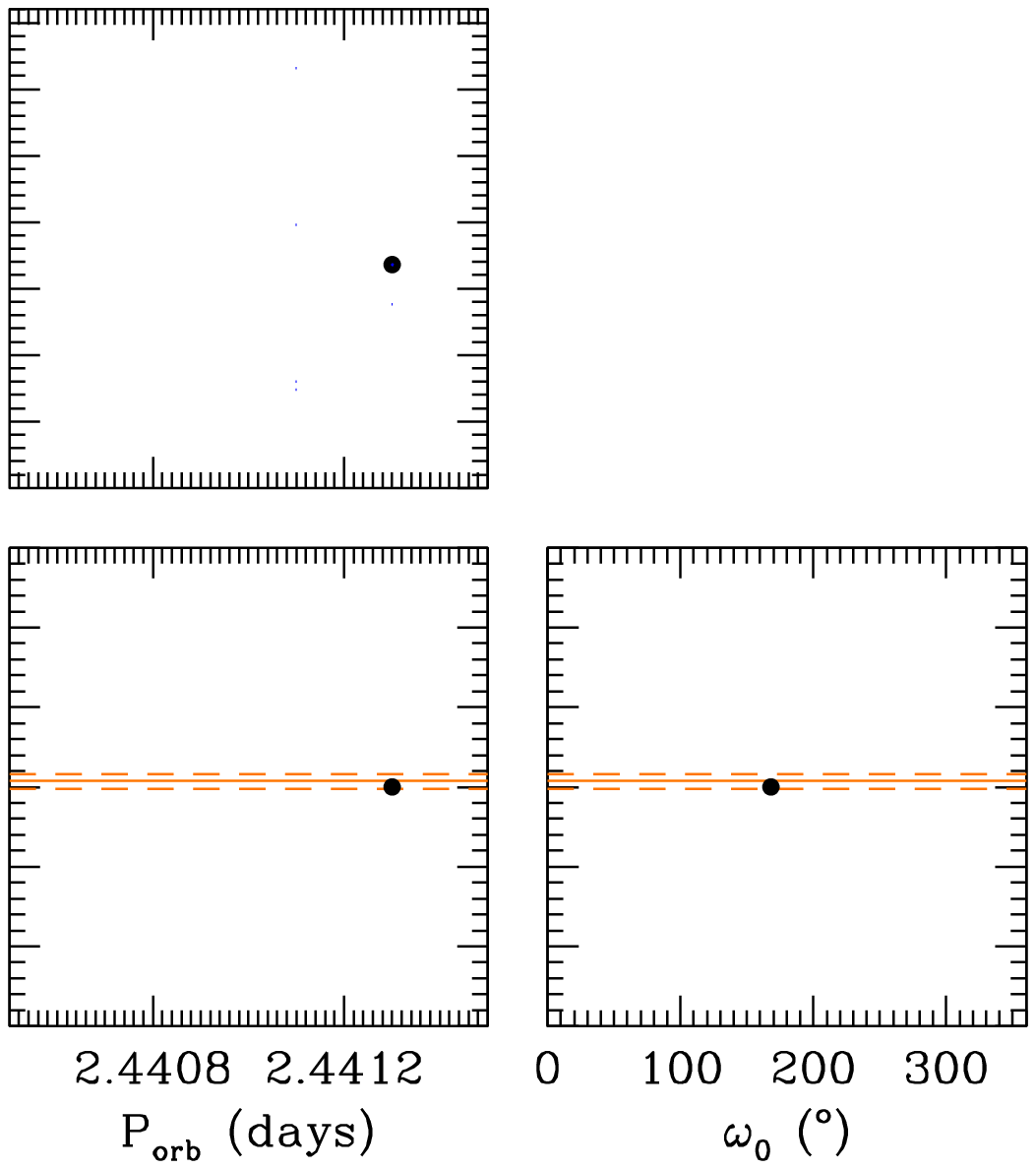}
\caption{Confidence contours for the best-fit parameters obtained from the adjustment of the primary RV data of CPD-41$^\circ$7742. The best-fit solution is shown in each panel by the black filled dot. The corresponding $1\,\sigma$ confidence level is shown by the blue contour. For comparison, the value of the apsidal motion rate (respectively eccentricity) and its error bars as derived in Sect.\,\ref{sect:times_minima} are shown in orange (respectively pink) respectively by the solid and dashed lines.  \label{fig:contours_RVs}}
\end{figure*}

To find the values of the six free parameters ($P_{\rm orb}$, the eccentricity $e$, the time $T_0$ of periastron passage, the argument of periastron of the primary orbit $\omega_0$ at epoch $T_0$, $\dot{\omega}$, and the semi-amplitude of the RV curve $K_{\rm P}$) that provide the best fit to the whole set of primary RV data, we scanned the parameter space in a systematic way. The projections of the 6-D parameter space onto the 2-D planes are illustrated in Fig.\,\ref{fig:contours_RVs}. We observe a strong degeneracy between the different parameters of the fit, as shown by the numerous local minima. We are therefore not able to derive a unique solution with sufficient accuracy, at least without the knowledge of one of these parameters.
 
\begin{figure}[h!]
\centering
\includegraphics[clip=true,trim=0cm 1cm 0cm 1.4cm, width=\linewidth]{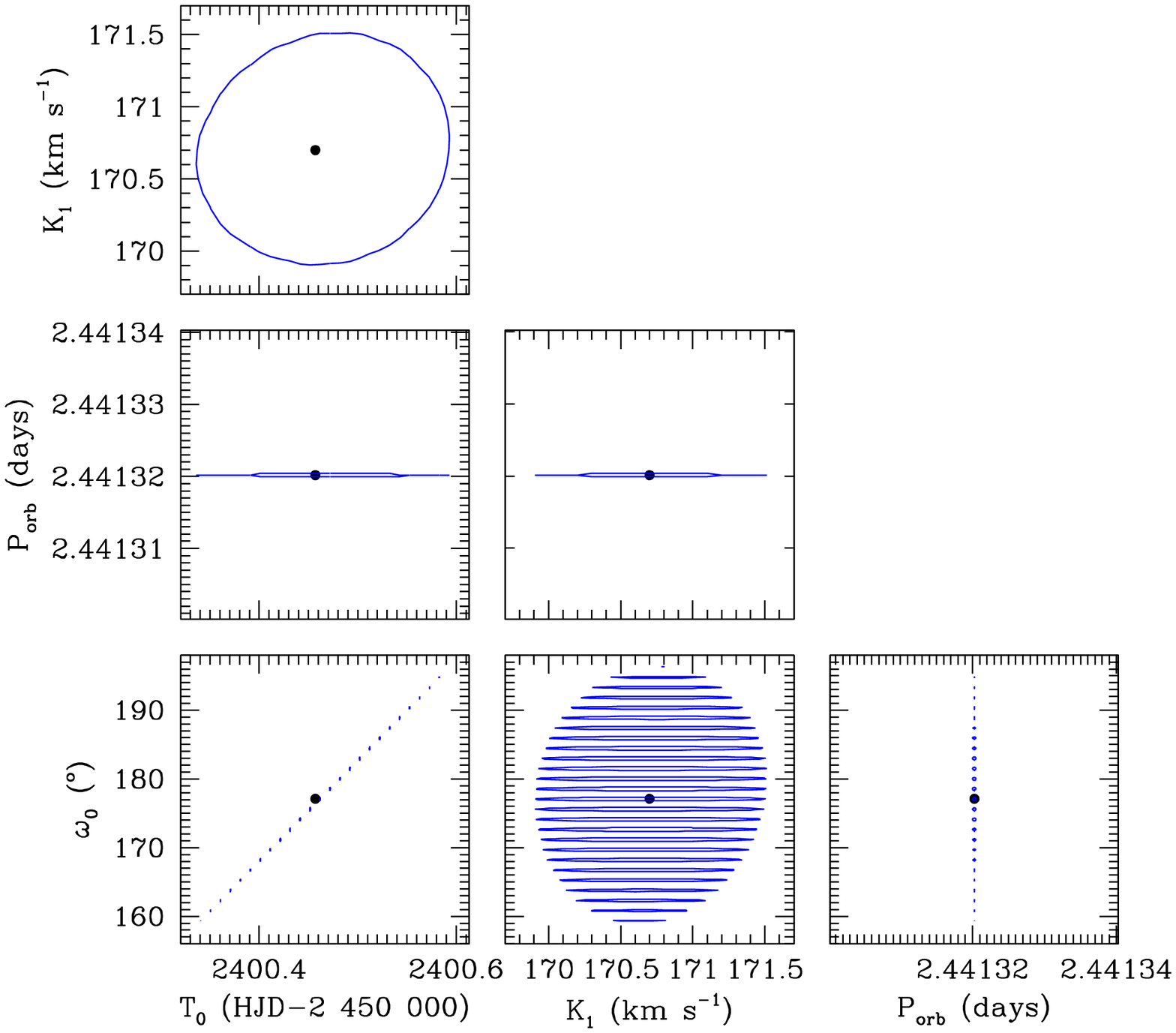}
\caption{Confidence contours for the best-fit parameters obtained from the adjustment of the primary RV data of CPD-41$^\circ$7742 when the eccentricity and the apsidal motion rate are fixed to the values of 0.0204 and $15\fdg38$\,yr$^{-1}$ as derived from the analysis of the times of photometric minima (see Sect.\,\ref{sect:times_minima}). The best-fit solution is shown in each panel by the black filled dot. The corresponding $1\,\sigma$ confidence level is shown by the blue contour.\label{fig:contours_efixed_omegadotfixed}}
\end{figure}

\begin{figure*}[h]
\centering
\includegraphics[width=\linewidth]{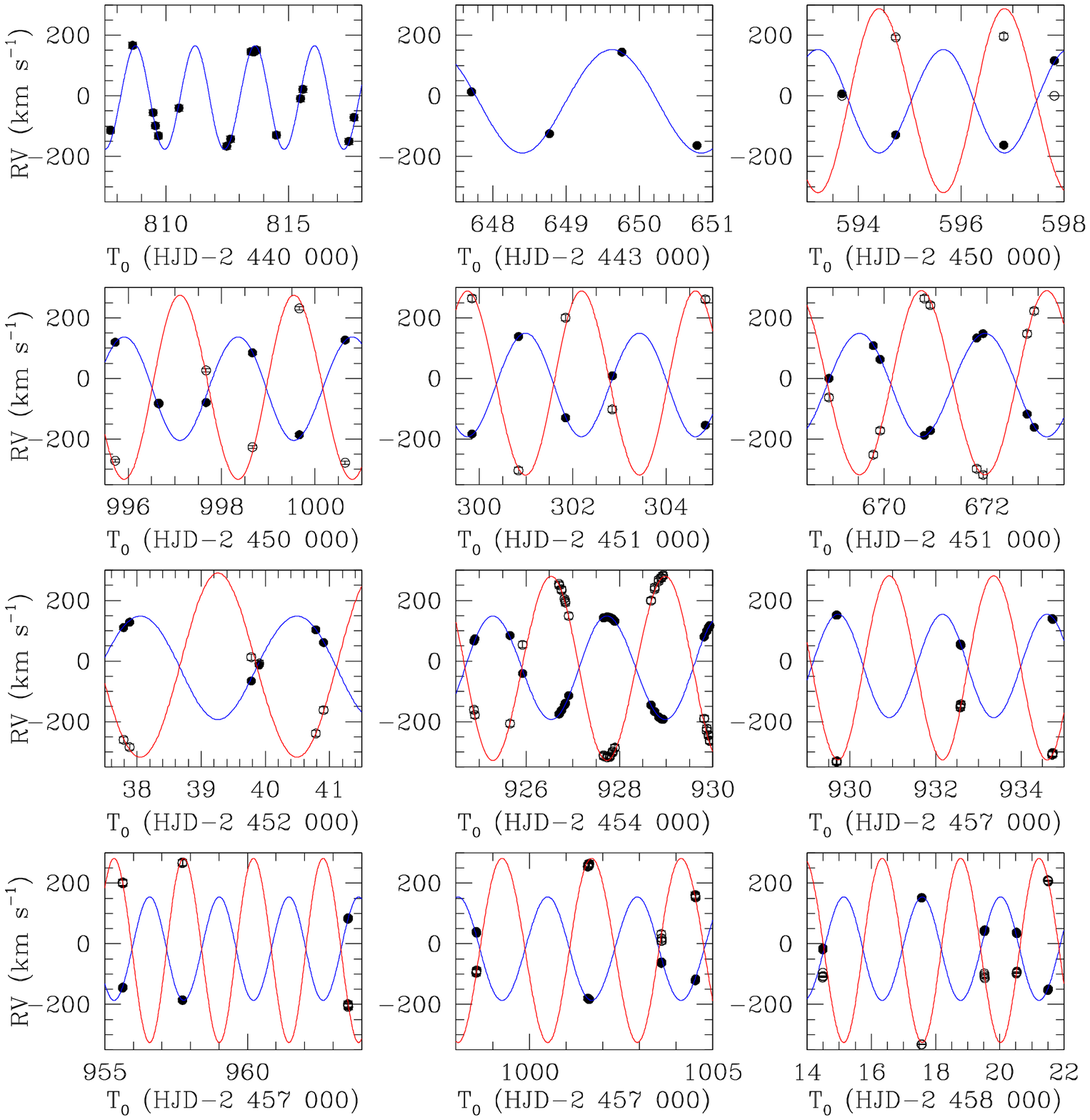}
\caption{Comparison between the measured RVs of the primary (filled dots) and secondary (open dots, when available) of CPD-41$^\circ$\,7742 with the orbital solution from Table\,\ref{bestfitTable}. The blue and red lines represent the fitted RV curve of the primary and secondary stars, respectively.  The top panels correspond to data from \citet[][\textit{left}]{Hil74}, \citet[][\textit{middle}]{Lev83}, and \citet[][\textit{right}]{Gar01}. The left panel on the second row yields the CAT/CES data from \citet{sana03}. All other panels correspond to RVs derived in this paper. \label{fitRV}}
\end{figure*}

To solve this issue, we fixed the eccentricity and apsidal motion rate to the values of 0.0204 and $15\fdg38$\,yr$^{-1}$ as derived from the analysis of the times of minimum (see Sect.\,\ref{sect:times_minima}), and left the four other parameters free. As Fig.\,\ref{fig:contours_RVs} indicates, this value of $\dot\omega$ is in agreement with one local minimum of the global adjustment of the RVs. The projections of the 4-D parameter space onto the 2-D planes are illustrated in Fig.\,\ref{fig:contours_efixed_omegadotfixed}. The corresponding orbital parameters are given in Table\,\ref{bestfitTable}. Figure\,\ref{fitRV} illustrates the best fit of the RV data at 12 different epochs, where the primary and secondary RV curves are depicted in blue and red, respectively. 

\begin{table}[h]
\caption{Best-fit orbital parameters of CPD-41$^\circ$7742 obtained from the adjustment of the RV data.}
\centering
\begin{tabular}{l l}
\hline\hline
\vspace{-3mm}\\
Parameter & Primary RVs \\
\hline
\vspace*{-3mm}\\
$e$ & 0.0204 (fixed) \\ 
\vspace*{-3mm}\\
$\dot{\omega}$ ($^{\circ}$\,yr$^{-1}$) & 15.38 (fixed) \\
\vspace*{-3mm}\\
$P_{\rm orb}$\,(d) &  $2.441320 \pm 0.000001$ \\
\vspace*{-3mm}\\
$\omega_0$  & $177\fdg1^{+19\fdg0}_{-18\fdg0}$\\
\vspace*{-3mm}\\
$T_0$ (HJD) & $2\,452\,400.46 \pm 0.12$ \\
\vspace*{-3mm}\\
$K_{\rm P}$\,(km\,s$^{-1}$) & $170.7\pm0.8$ \\
\vspace*{-3mm}\\
$\chi^2_\nu$ & 2.028 \\
\vspace*{-3mm}\\
\hline
\vspace*{-3mm}\\
$q = M_\text{S}/M_\text{P}$   & $0.562 \pm 0.006$ \\
\vspace*{-3mm}\\
$K_{\rm S}$\,(km\,s$^{-1}$) &  $303.7\pm 3.5$ \\
\vspace*{-3mm}\\
$a_{\rm P}\,\sin{i}$\,($R_{\odot}$) & $8.23 \pm 0.06$ \\
\vspace*{-3mm}\\
$a_{\rm S}\,\sin{i}$\,($R_{\odot}$) & $14.64 \pm 0.12$\\
\vspace*{-3mm}\\
$M_{\rm P}\,\sin^3{i}$\,($M_{\odot}$) & $17.27 \pm 0.49$ \\
\vspace*{-3mm}\\
$M_{\rm S}\,\sin^3{i}$\,($M_{\odot}$) & $9.71 \pm 0.29$ \\
\vspace*{-3mm}\\
\hline
\end{tabular}
\label{bestfitTable}
\end{table} 

The primary and secondary RVs of an SB2 binary are related to each other through the mass ratio $q=\frac{M_{\rm S}}{M_{\rm P}}$. We derived a value $q =0.562 \pm 0.006$ from the RVs coming from the disentangling process, and we used this result to build an SB2 orbital solution for CPD-41$^\circ$\,7742. Our best-fit parameters and their $1\sigma$ errors are listed in Table\,\ref{bestfitTable}, where $a_\text{P}\sin i$ and $a_\text{S}\sin i$ stand for the minimum semi-major axis of the primary and secondary stars, respectively, $M_\text{P}\sin^3 i$ and $M_\text{S}\sin^3 i$ stand for the minimum mass of the primary and secondary stars, respectively, and $\chi^2_\nu$ is the reduced $\chi^2$.

We note that the mass ratio we found is slightly higher than the value of $0.555 \pm 0.005$ quoted by \citet{sana03} from the He\,{\sc i} lines but is compatible with this value within the error bars. We further note that our semi-amplitudes of the primary and secondary RV curves are higher than those derived by \citet[see Table\,\ref{table:intro}]{sana03}. These differences are likely due to the two-Gaussian fit used by \citet{sana03} to establish their RVs. 

\section{Photometric analysis \label{sect:photom}}
The \texttt{Nightfall} binary star code (version 1.92), developed by R. Wichmann, M. Kuster, and P. Risse\footnote{The code is available at:\\http://www.hs.uni-hamburg.de/DE/Ins/Per/Wichmann/Nightfall.html} \citep{Wichmann}, was used to analyse the light curves of the binary CPD-41$^\circ$\,7742. The Roche potential, scaled with the instantaneous separation between the stars, is used to describe the stellar shape. We neglected any stellar surface spots. In total, we have eight model parameters: the effective temperatures $T_\text{eff,P}$ and $T_\text{eff,S}$ of the primary and secondary stars, the Roche lobe filling factors (ratios between the stellar polar radius and the polar radius of the associated instantaneous Roche lobe at periastron) $f_\text{P}$ and $f_\text{S}$ of the primary and secondary stars, the mass ratio $q$, the orbital eccentricity $e$, the orbital inclination $i$, and the argument of periastron $\omega$. As the Bochum and $uvby$ observations are not polluted by the presence of another star in the field, there is no need to include any third light in the model. The TESS observations are, however, polluted by the presence of the massive binary HD\,152248 in the field, and hence, a third light contribution is included in the model for these observations. Such a third light contribution leads to further degeneracies among the best-fit solutions of the light curve. Hence, we decided to analyse the TESS data with the stellar parameters set to their values inferred from the fit of the Bochum data.

The period of the system was fixed to the instantaneous sidereal period $P_\text{ecl}$, that is to say, the time interval between two primary minima. Its expression as a function of the orbital period, the eccentricity, the apsidal motion rate, and the argument of periastron is given by Eq.\,(A.24) of \citet{schmitt16}: 
\begin{equation}
\label{eqn:Pecl}
P_\text{ecl}=\left(1-\frac{(1-e^2)^{3/2}}{1+e\sin\omega}\frac{\dot\omega P_\text{orb}}{2\pi}\right)P_\text{orb}\,,
\end{equation}
where $\dot\omega$ is expressed in radians per day. For the value of $P_\text{orb}$ given in Table\,\ref{bestfitTable} and the values of $e$ and $\dot\omega$ given in Table\,\ref{table:times_minima}, $P_\text{ecl}$ ranges between 2.440609 and 2.440637\,d. 

The existence of apsidal motion in the system and the different levels of third light contamination of the various datasets prevent us from performing a simultaneous analysis of all photometric data. Hence, we performed the analysis of the four datasets (Bochum 6051 and 4686, $uvby$, TESS-12, and TESS-39) separately. The dates of primary minima were directly adjusted on the phase-folded light curve. We obtained 2\,450\,553.544, 2\,451\,932.48551, 2\,458\,629.5717, and 2\,459\,361.7536 HJD for the Bochum, $uvby$, TESS-12, and TESS-39 photometry, respectively.

The photometric data collected with the Bochum telescope cover an interval of 28\,days, which is short enough to neglect any change of $\omega$. We fixed the effective temperatures of the stars to the values derived from the \texttt{CMFGEN} analysis (see Table\,\ref{Table:CMFGEN} and Sect.\,\ref{subsect:cmfgen}). We fixed the mass ratio to the value derived from the RV analysis (see Table\,\ref{bestfitTable}). We were thus left with five parameters to adjust: $i$, $e$, $f_\text{P}$, $f_\text{S}$, and $\omega$. 
In this way, the best-fit Roche-lobe filling factor of the primary star (near 0.73) was systematically smaller than that of the secondary star ($\simeq 0.83$), leading to $f_\text{S}/f_\text{P} = 1.14$. With the stellar temperatures fixed to their values derived from the spectroscopic analysis, this result is at odds with the spectroscopic brightness ratio that rather implies $R_\text{P}/R_\text{S} = 2.05 \pm 0.45$ and thus $f_\text{S}/f_\text{P} = 0.63 \pm 0.14$.

\begin{figure}[h]
\includegraphics[clip=true, trim=20 40 30 40,width=\linewidth]{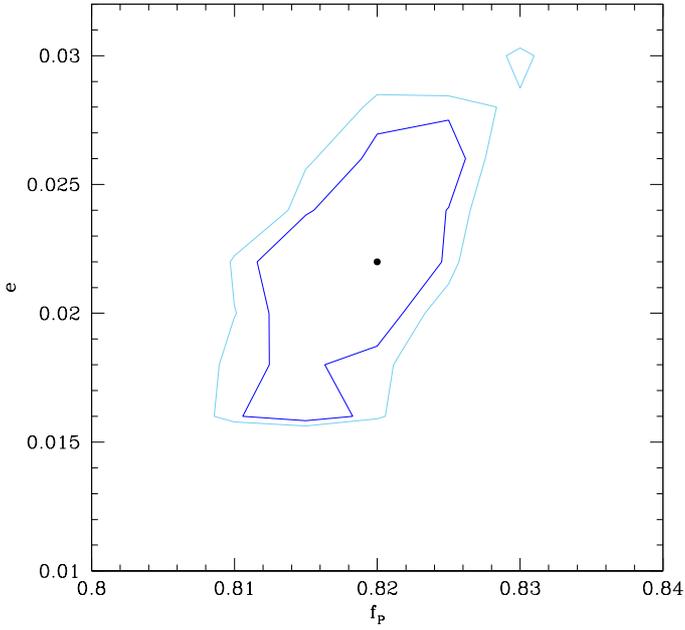}
\caption{Confidence contours of the best-fit solutions of the Bochum photometry of CPD-41$^\circ$\,7742 in the $(f_\text{P}, e)$-plane. The best-fit solution is shown by the black filled dot. The corresponding $1\sigma$  and 90\% confidence levels are shown by the dark and light blue contours, respectively.\label{contNFBochum}}
\end{figure}

\begin{figure}[p]
\centering
\includegraphics[clip=true,trim=60 60 40 200,width=0.925\linewidth]{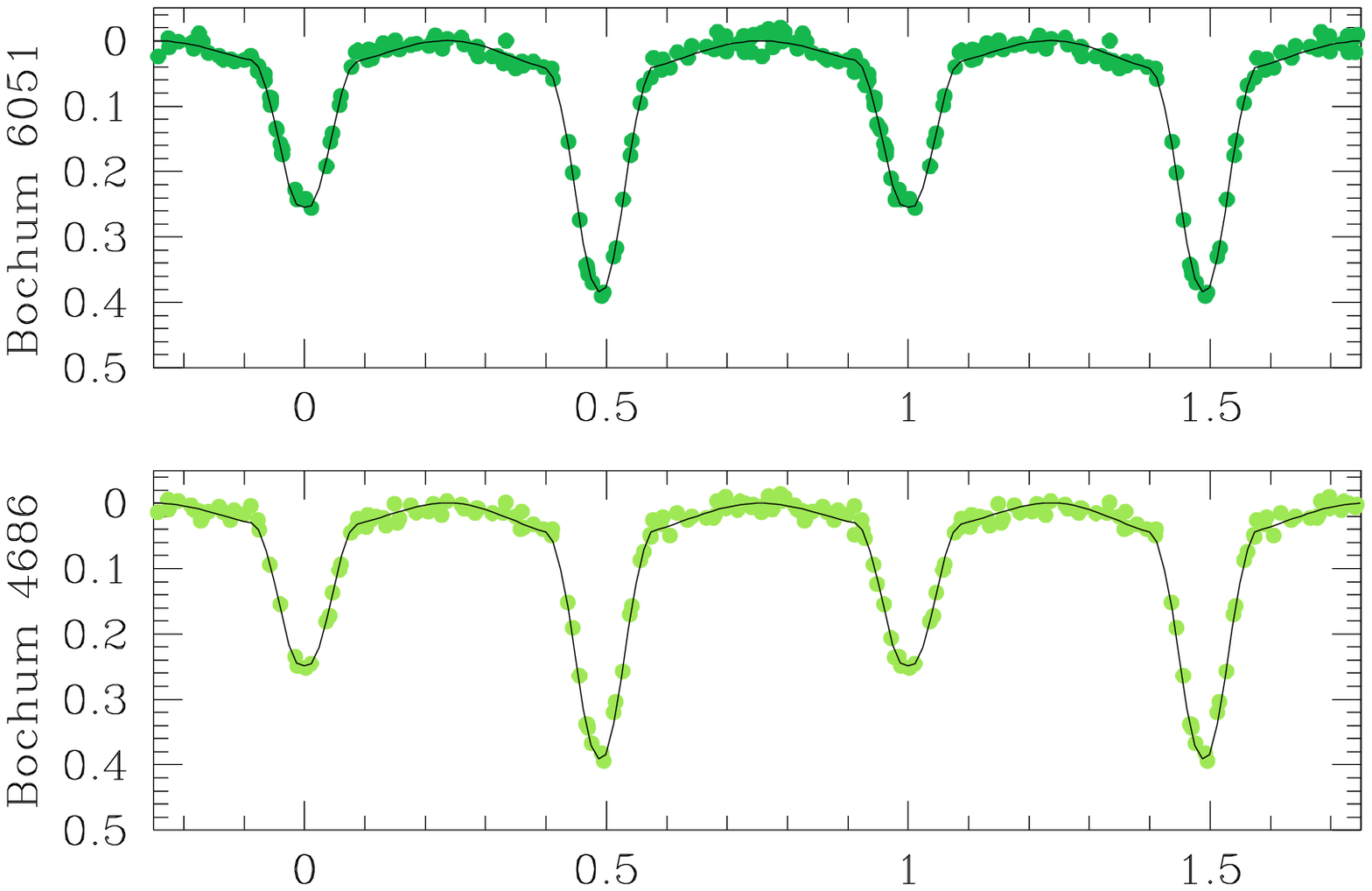}\\
\includegraphics[clip=true,trim=60 60 40 200,width=0.925\linewidth]{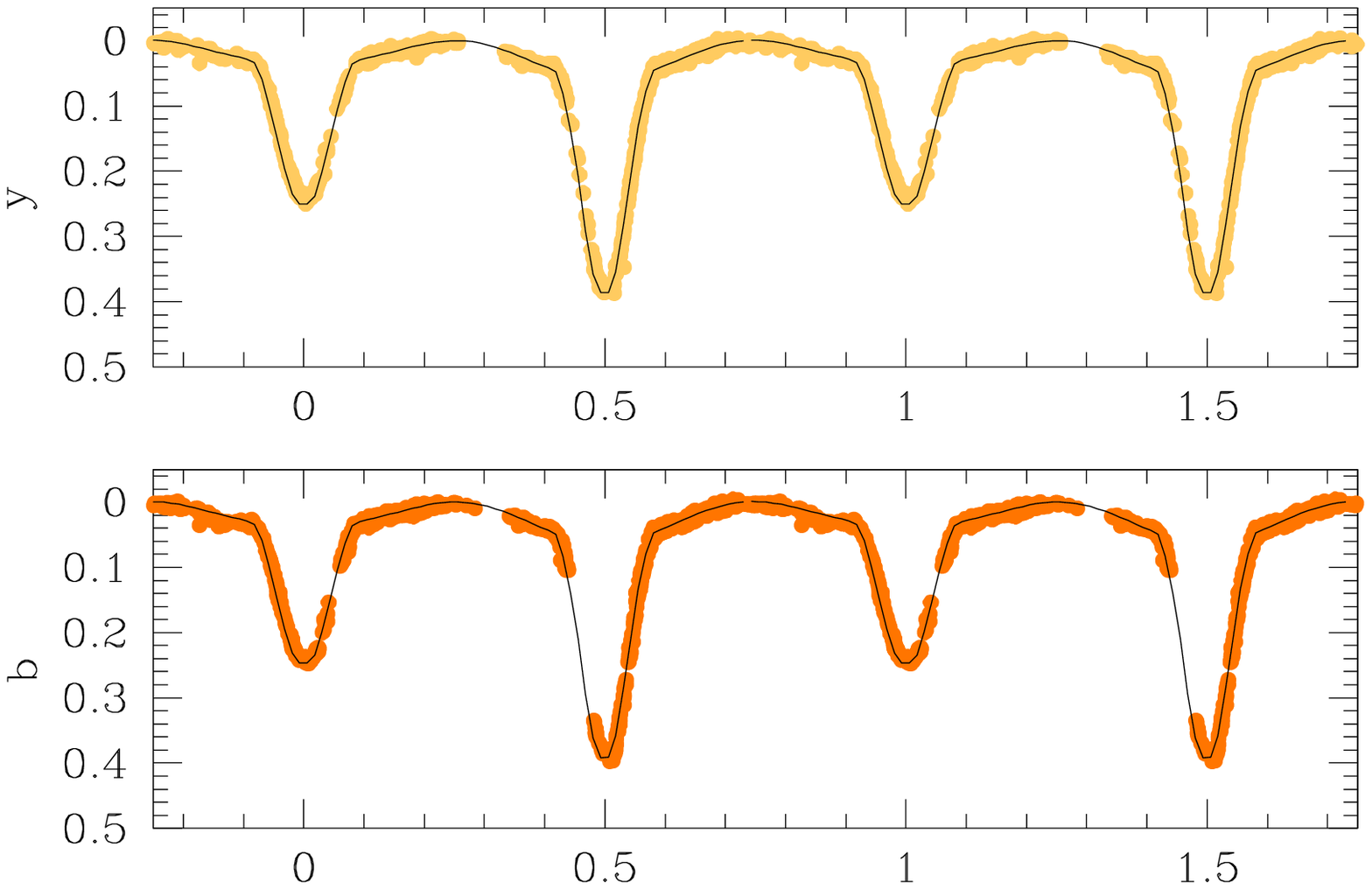}\\
\includegraphics[clip=true,trim=60 60 40 200,width=0.925\linewidth]{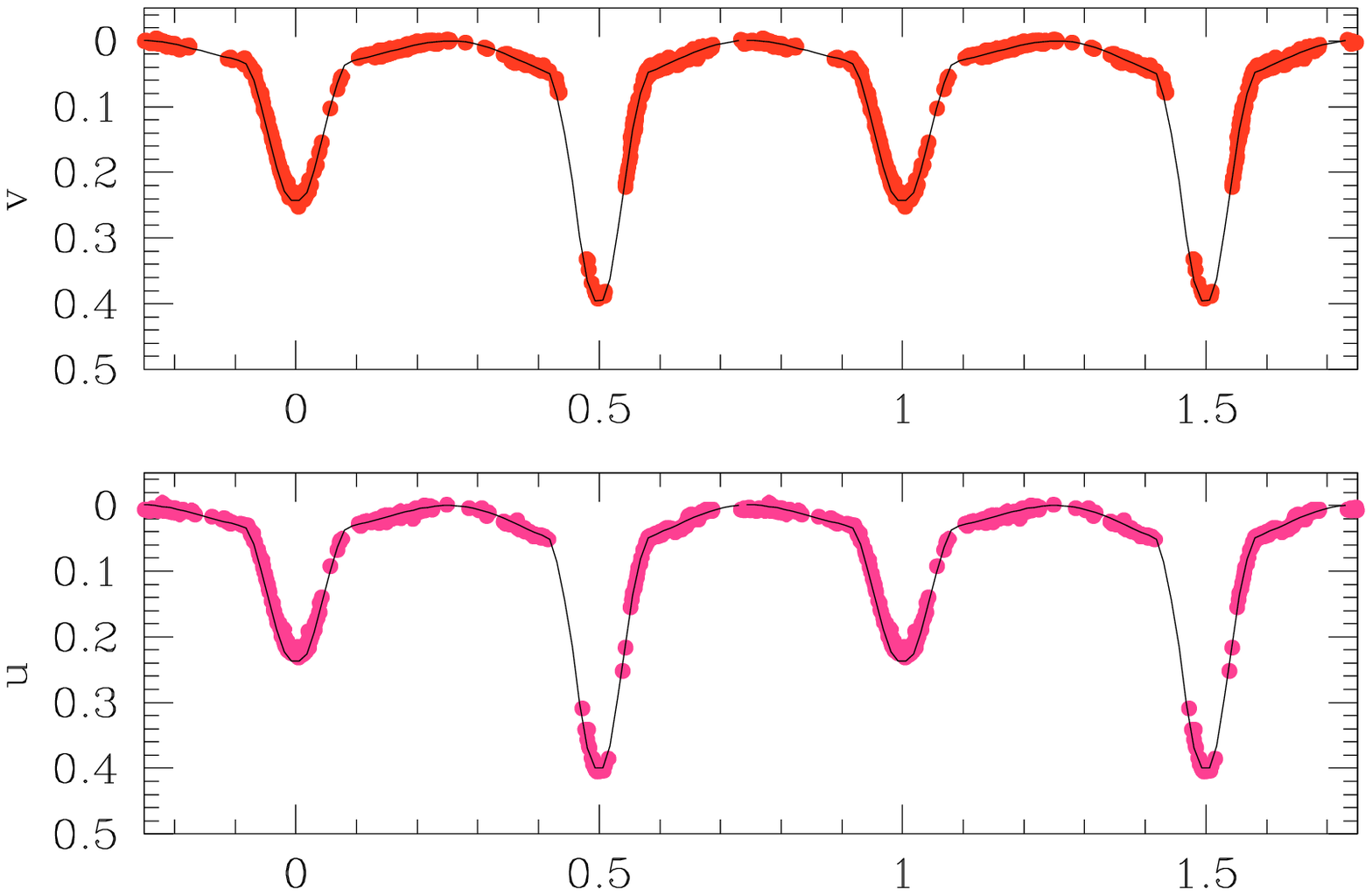}\\
\includegraphics[clip=true,trim=60 40 40 200,width=0.925\linewidth]{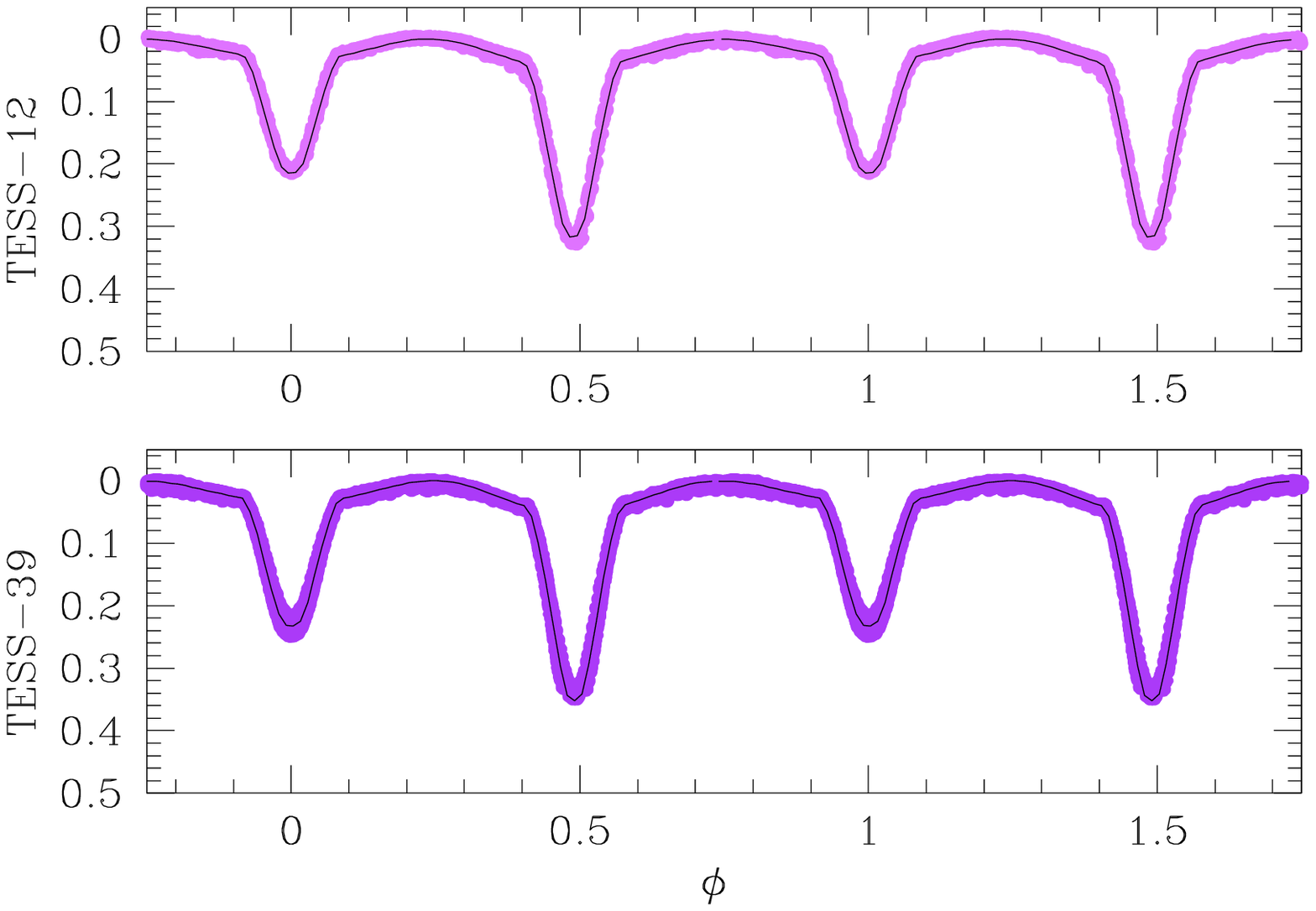}
 \caption{Best-fit solution to the Bochum 6051 and 4686, $uvby$, TESS-12, and TESS-39 photometric data of CPD-41$^\circ$\,7742. The light curves have been phase-folded using $P_\text{ecl}$. The dates of primary minima are 2\,450\,553.544, 2\,451\,932.48551, 2\,458\,629.5717, and 2\,459\,361.7536 HJD for the Bochum, $uvby$, TESS-12, and TESS-39 photometry, respectively.
 \label{fig:photom}}
\end{figure}

To solve this issue, we explored a grid of models where we varied the value of $f_\text{P}$ between 0.80 and 0.90 and fixed $f_\text{S}$ in such a way that the ratio $f_\text{S}/f_\text{P}$ is equal to the spectroscopic value or this value $\pm 1\,\sigma$. These tests indicated that the best-fit solutions were obtained when $f_\text{S}$ was at the higher end of the `authorised' range and for $f_\text{P}$ values near $0.81$. To account for the fact that the spectroscopic temperatures are subject to some uncertainties, we then started exploring another grid of models for $f_\text{P}$ between 0.80 and 0.84, but this time leaving $f_\text{S}$ and the secondary's temperature as free parameters. The main results of this approach are summarised in Table\,\ref{resfitsBochum}. The $1\sigma$ confidence contour in the $(f_\text{P}, e)$-plane is shown in Fig.\,\ref{contNFBochum} and the best fit is shown in Fig.\,\ref{fig:photom}. The eclipse depths of both eclipses of the two Bochum datasets are given in Table\,\ref{table:eclipse_depths}.

\begin{table}[h!]
\centering
\caption{Best-fit parameters of the Bochum 6051 and 4686 photometric data of CPD-41$^\circ$\,7742. \label{resfitsBochum}}
\begin{tabular}{l l}
\hline
\hline
\vspace*{-3mm}\\
Parameter & Value \\
\hline
\vspace*{-3mm}\\
$q$ & 0.562 (fixed) \\
\vspace*{-3mm}\\
$T_{\text{eff,P}}$ (K) & 31\,800 (fixed) \\
\vspace*{-3mm}\\
$T_{\text{eff,S}}$ (K) & $24\,098^{+50}_{-100}$ \\
\vspace*{-3mm}\\
$f_\text{P}$ & $0.820^{+0.007}_{-0.008}$ \\
\vspace*{-3mm}\\
$f_\text{S}$ & $0.604^{+0.003}_{-0.007}$ \\
\vspace*{-3mm}\\
$i$  & $82\fdg0^{+0.6}_{-0.5}$ \\
\vspace*{-3mm}\\
$e$ & $0.022^{+0.005}_{-0.006}$ \\
\vspace*{-3mm}\\
$\omega$  & $40\fdg4^{+14.2}_{-28.0}$ \\
\vspace*{-3mm}\\
$\chi^2_\nu$ &1.87 \\
\vspace*{-3mm}\\
\hline
\end{tabular}
\end{table}

\begin{table}[h]
\centering
\caption{Time at mid-exposure and depths of the primary and secondary eclipses for the photometric observations of CPD-41$^\circ$\,7742.  \label{table:eclipse_depths}}
\begin{tabular}{l l l l}
\hline
\hline
\vspace*{-3mm}\\
Filter  & Time (HJD) & \multicolumn{2}{c}{Eclipse depth (mag)}\\
& & Primary & Secondary \\
\hline
\vspace*{-3mm}\\
Bochum 6051 & 2\,450\,544.8142&  0.255&0.384 \\
\vspace*{-3mm}\\
Bochum 4686 &2\,450\,546.8402&0.248 &0.391 \\
\vspace*{-3mm}\\
$u$-band &2\,451\,911.7223& 0.237& 0.400\\
\vspace*{-3mm}\\
$v$-band &2\,451\,911.7094& 0.243& 0.396\\
\vspace*{-3mm}\\
$b$-band &2\,451\,911.7205& 0.246& 0.391\\
\vspace*{-3mm}\\
$y$-band & 2\,451\,911.7251 & 0.251& 0.386\\
\vspace*{-3mm}\\
TESS-12 &2\,458\,646.4641 & ... & ... \\
\vspace*{-3mm}\\
TESS-39 & 2\,459\,373.9463 & ... & ... \\
\vspace*{-3mm}\\
\hline
\end{tabular}
\tablefoot{ We do not provide the eclipse depths of the TESS-12 and TESS-39 data as the TESS light curves are diluted by a residual contamination of third light.}
\end{table}

Compared to solutions with the secondary effective temperature fixed, the quality of the fits is improved. The secondary star's best-fit photometric temperature is lower than the spectroscopic value by about 2000\,K (i.e.\ twice the estimated error on the spectroscopic value). Yet, it should be noted that the primary spectroscopic temperature also has an uncertainty of 1000\,K, which is not accounted for in the photometric fits. Since the photometric solution is mostly sensitive to the ratio of the temperatures, increasing the primary temperature by 1000\,K (i.e.\ $1\sigma$ of the spectroscopic value) actually leads to a secondary effective temperature of 24\,960\,K, which is at a bit more than $1\sigma$ from the spectroscopic value. For consistency, in the following, we adopt an error of 1000\,K on the secondary effective temperature.
The $V$-band brightness ratio of the two stars for these solutions amounts to $\simeq 0.20$ in excellent agreement with the spectroscopic value. Hence, the issue mentioned above can be solved in this way.
The value of 0.0204 for the eccentricity derived from the analysis of the times of minima (see Sect.\,\ref{sect:times_minima}) is fully consistent with the best estimate obtained here. 

Finally, we analysed the $uvby$, TESS-12, and TESS-39 photometry fixing all parameters to their best-fit values (see Table\,\ref{resfitsBochum}) except for $e$ fixed to the value of 0.0204 (see Sect.\,\ref{sect:times_minima}), and for $\omega$ that was left as a free parameter. For the $uvby$ photometry, the best fit gives $\omega = 84\fdg6^{+0.7}_{-0.5}$ and is illustrated in Fig.\,\ref{fig:photom}. The four light curves are well-adjusted. In order to check if our best-fit solution is robust and not biased by the best-fit solution of the Bochum photometry, we performed three additional adjustments of the $uvby$ photometry: For all three cases, in addition to $\omega$, the secondary effective temperature was left free; for two of these, the secondary Roche lobe filling factor was also left free; and for one of these, the inclination was left as a free parameter too. The results are given in Table\,\ref{tab:bestfit:uvby}. The reduced $\chi^2_\nu$ decreases slightly as more free parameters are adopted, but the best-fit values of the parameters do not change significantly, which ensures that our best-fit solution with all parameters fixed except for $\omega$ is robust.  The eclipse depths of both eclipses of the photometric datasets are given in Table\,\ref{table:eclipse_depths}. Regarding the TESS-12 and TESS-39 data, we also left the third light contribution $I_3$ as a free parameter of the adjustment and obtained $\omega=347\fdg0^{+3.0}_{-5.0}$, $I_3=0.147\pm 0.010$, and $\chi^2_\nu=694.4$, and $\omega=43\fdg4^{+6.6}_{-3.4}$, $I_3=0.071\pm0.012$, and $\chi^2_\nu=347.0$, respectively. We note that the large values of $\chi_\nu^2$ have no physical sense and arise because of the underestimate of the errors on the TESS data. The best fits are shown in Fig.\,\ref{fig:photom}. As the eclipses of the TESS light curves are diluted by the residual contamination of third light, we cannot compare their depths to those of the Bochum and $uvby$ photometric data. 

\begin{table}[h]
\centering
\tiny
\caption{Best-fit parameters of the $uvby$ photometry of CPD-41$^\circ$\,7742. \label{tab:bestfit:uvby}}
\begin{tabular}{l l l l l}
\hline\hline
\vspace*{-3mm}\\
Parameter & \multicolumn{4}{c}{Value} \\
& Case I & Case II & Case III & Case IV \\
\hline
\vspace*{-3mm}\\
$i$  & $82\fdg0$ (fixed) & $82\fdg0$ (fixed) & $82\fdg0$ (fixed) & $80\fdg8^{+0.9}_{-0.4}$  \\
\vspace*{-3mm}\\
$f_\text{S}$ & $0.604$ (fixed) &$0.604$ (fixed)  & $0.593^{+0.003}_{-0.001}$ & $0.613^{+0.012}_{-0.015}$ \\
\vspace*{-3mm}\\
$T_{\text{eff,S}}$ (K) & $24\,098$ (fixed) & $23\,408^{+110}_{-100}$ & $23\,726^{+50}_{-200}$ & $23\,700^{+200}_{-160}$ \\
\vspace*{-3mm}\\
$\omega$  & $84\fdg6^{+0.7}_{-0.5}$ & $84^{+0.7}_{-1.1}$ & $85\fdg8^{+1.5}_{-2.3}$ & $87\fdg7_{-2.3}^{+1.7}$ \\
\vspace*{-3mm}\\
$\chi^2_\nu$ & 0.857 & 0.638 & 0.556 & 0.534 \\
\vspace*{-3mm}\\
\hline
\end{tabular}
\end{table}

\section{Times of minimum\label{sect:times_minima}}
The times of minimum of photometric observations can serve for the determination of the apsidal motion rate of an eclipsing binary system \citep{gimenez95}. We computed the phase difference between the primary and secondary eclipses, 
\begin{equation}
\label{eqn:delta_phi}
\Delta\phi = \frac{T_2-T_1}{P_\text{ecl}},
\end{equation}
for the different sets of observations. We adjusted a second-order polynomial as well as a Gaussian to the eclipses for the Bochum 6051 and 4686, $uvby$, TESS-12, and TESS-39 separately, and found the values given in Table\,\ref{table:times_minima_obs}. 

\begin{table}[h]
\centering
\caption{Phase difference given by Eq.\,\eqref{eqn:delta_phi} between the primary and secondary eclipses for the photometric observations of CPD-41$^\circ$\,7742.\label{table:times_minima_obs}}
\begin{tabular}{l l}
\hline\hline
\vspace*{-3mm}\\
Filter & $\Delta\phi$ \\
\hline
\vspace*{-3mm}\\
Bochum 6051 \& 4686 & $0.5109 \pm 0.0010$ \\
\vspace*{-3mm}\\
$uvby$-bands & $0.5011 \pm 0.0008$\\
\vspace*{-3mm}\\
TESS-12 & $0.5132 \pm 0.0007$   \\
\vspace*{-3mm}\\
TESS-39 & $0.5099 \pm 0.0007$ \\
\vspace*{-3mm}\\
\hline
\end{tabular}
\end{table}

\begin{figure}[]
\centering
\includegraphics[clip=true,trim=30 40 10 130,width=\linewidth]{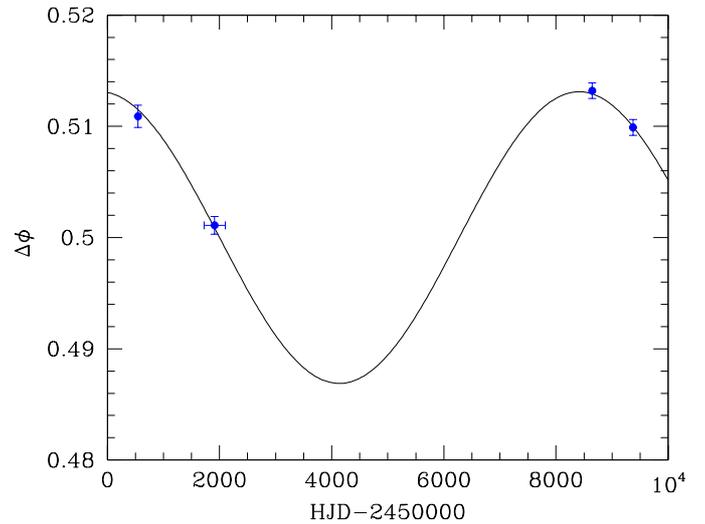}
\caption{Values of the phase difference $\Delta\phi$ between the primary and secondary eclipses as a function of time inferred from the times of minimum of the light curves. The blue symbols correspond to the data of the fits of the Bochum 6051 and 4686, $uvby$, TESS-12, and TESS-39 photometry.  The solid line corresponds to our best-fit value of $\Delta\phi$ from Eq.\,\eqref{eqn:times_minima}. \label{fig:timing}}
\end{figure}

\begin{figure}[]
\centering
\includegraphics[clip=true,trim=40 30 40 60,width=1\linewidth]{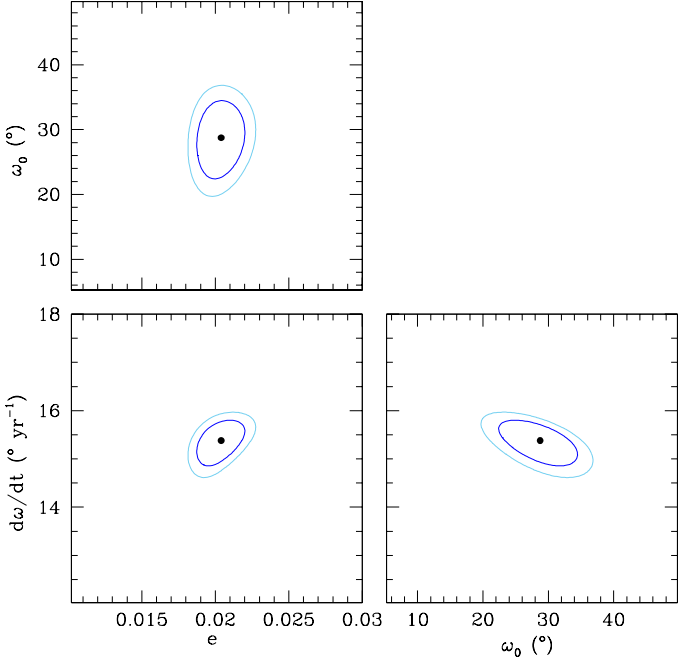}
\caption{Confidence contours for the best fit obtained from the adjustment of the phase differences between the times of the secondary and primary eclipses using Eq.\,\eqref{eqn:times_minima}. The best-fit solution is shown in each panel by the black filled dot. The corresponding $1\sigma$ and 99\% confidence levels are shown by the dark and light blue contours, respectively.  \label{fig:times_minima_contours}}
\end{figure}

We then computed the phase difference between the times of the secondary and primary eclipses, respectively $T_2$ and $T_1$, following the relation adapted from \citet{gimenez95}: 
\begin{equation}
\label{eqn:times_minima}
\Delta\phi =  \frac{1}{2} + A_1 \frac{e}{\pi} \cos(\omega) - A_3 \frac{e^3}{4\pi} \cos(3\omega) + A_5 \frac{e^5}{16\pi}\cos(5\omega), 
\end{equation}
where $A_1$, $A_3$, and $A_5$ are functions of the inclination and the eccentricity given by Eqs.\,(16), (18), and (20) of \citet{gimenez95}. We adjusted this curve to the observations explicitly accounting for the apsidal motion rate $\dot\omega$ through the variation of $\omega$ with time. This is illustrated in Fig.\,\ref{fig:timing}. The best-fit adjustment (summarised in Table\,\ref{table:times_minima}) is obtained for $e=0.0204 \pm 0.0016$, $\dot\omega=(15\fdg38^{+0.42}_{-0.51})\,\text{yr}^{-1}$, $\omega=28\fdg75^{+5.63}_{-6.38}$ at the time of the Bochum observations, and $\chi_\nu=0.1539$. The projections of the 3-D parameter space onto the 2-D planes are illustrated in Fig.\,\ref{fig:times_minima_contours}. 

\begin{table}[]
\centering
\caption{Best-fit parameters of the adjustment of the phase difference between the primary and secondary eclipses for the photometric observations of CPD-41$^\circ$\,7742.\label{table:times_minima}}
\begin{tabular}{l l}
\hline\hline
\vspace*{-3mm}\\
Parameter & Value\\
\hline
\vspace*{-3mm}\\
$e$ & $0.0204 \pm 0.0016$ \\
\vspace*{-3mm}\\
$\dot\omega$ ($^\circ\,\text{yr}^{-1}$)& $15.38^{+0.42}_{-0.51}$\\
\vspace*{-3mm}\\
$T_0$ (HJD -- 2\,450\,000) & 553.544 \\
\vspace*{-3mm}\\
$\omega_0$  & $28\fdg75^{+5.63}_{-6.38}$\\
\vspace*{-3mm}\\
$\chi_\nu^2$ &  0.1539\\
\vspace*{-3mm}\\
\hline
\end{tabular}
\end{table}

\section{CPD-41$^\circ$\,7742 orbital and physical parameters\label{sect:summary}}
The best-fit parameters obtained from the adjustment of the RVs (see Table\,\ref{bestfitTable}), the Bochum light curves (see Table\,\ref{resfitsBochum}), and the adjustment of the times of minimum (see Table\,\ref{table:times_minima}) are combined to finally establish the orbital and physical parameters of CPD-41$^\circ$\,7742. A semi-major axis of $23.09\pm0.16\,R_\odot$ is deduced for the system. Absolute masses for the primary and secondary stars $M_\text{P} = 17.8 \pm 0.5\,M_\odot$ and $M_\text{S}=10.0\pm 0.3\,M_\odot$, as well as absolute radii for the primary and secondary stars $R_\text{P} = 7.57 \pm 0.09\,R_\odot$ and $R_\text{S}=4.29^{+0.04}_{-0.06}\,R_\odot$ are derived. This leads to photometric values of the surface gravities of the primary and secondary stars $\log g_\text{P} = 3.93\pm0.02$ and $\log g_\text{S} = 4.17\pm0.02$. 
From the photometric stellar radii, the primary effective temperature derived in Sect.\,\ref{subsect:cmfgen}, and the secondary effective temperature derived in Sect.\,\ref{sect:photom}, we inferred bolometric luminosities $L_{\text{bol,P}} = (5.28^{+0.67}_{-0.68})\times 10^4\,L_\odot$ and $L_{\text{bol,S}} = 5.58^{+0.93}_{-0.94}\times 10^3\,L_\odot$ for the primary and secondary stars, respectively.  
If we further assume the stellar rotational axes to be aligned with the normal to the orbital plane, the combination of the orbital inclination, the stellar radii, and the projected rotational velocities of the stars derived in Sect.\,\ref{subsect:vsini} yields rotational periods for the primary and secondary stars $P_{\text{rot,P}} = 2.71\pm 0.14$\,days and $P_{\text{rot,S}} = 2.42\pm 0.19$\,days. The ratio between rotational angular velocity and instantaneous orbital angular velocity at periastron amounts to $0.86 \pm 0.04$ and $0.97 \pm 0.08$ for the primary and secondary stars, respectively. Pseudo-synchronisation is achieved for the secondary star and about to be achieved for the primary star. \\

The apsidal motion rate of a binary system is the sum of a Newtonian contribution, $\dot\omega_\text{N}$, and a general relativity correction, $\dot\omega_\text{GR}$. The Newtonian contribution and the general relativistic correction amount to $14\fdg97^{+0.42}_{-0.51}$\,yr$^{-1}$ and $0\fdg41\pm0\fdg01\,\text{yr}^{-1}$, respectively, in the case of CPD-41$^\circ$\,7742. We refer to \citet{rosu22} for a review of the useful equations. The Newtonian contribution depends upon the internal mass distribution of the stars through the internal structure constants $k_2$ of the stars: 
\begin{equation}
\dot\omega_\text{N} = c_1 k_{2,1} + c_2 k_{2,2},
\end{equation}
where $c_1$ and $c_2$ are two functions of stellar and orbital parameters \citep{rosu22}.  We define a weighted-average mean of the internal structure constants, 
\begin{equation}
\bar{k}_2 = \frac{c_1 k_{2,1}+c_2 k_{2,2}}{c_1 + c_2} = \frac{\dot\omega_\text{N}}{c_1+ c_2},
\end{equation}
the value of which amounts to $0.0064^{+0.0004}_{-0.0005}$ in the present case. As the secondary star is smaller and less massive than the primary star, its $k_2$-value exceeds that of the primary. Therefore, we have $k_{2,1}< \bar{k}_2 < k_{2,2}$. Observationally, we have $\frac{c_1}{c_1+c_2} = 0.85$ and $\frac{c_2}{c_1+c_2}=0.15$, meaning that the $k_{2,1}$ has a much higher weight in the calculation of $\bar{k}_2$.

\section{Stellar structure and evolution models\label{sect:cles}}
We computed stellar evolution models with the Code Li\'egeois d'\'Evolution Stellaire\footnote{The \texttt{Cl\'es} code is developed and maintained by Richard Scuflaire at the STAR Institute at the University of Li\`ege.} \citep[\texttt{Cl\'es,}][see also \citet{rosu22} for an overview of the main features of \texttt{Cl\'es} and its minimisation routine \texttt{min-Cl\'es} used in the present context]{scuflaire08}. The aim of the present analysis is to see whether standard models reproducing the stellar physical properties are also capable of reproducing the apsidal motion rate inferred in Sect.\,\ref{sect:times_minima}. For all models built, we adopted the mass-loss scaling factor $\xi = 1$.  \\

In a first attempt to obtain stellar evolution models representative of both stars, we built two \texttt{Cl\'es} models (Models I$_\text{P}$ and I$_\text{S}$) simultaneously, adopting the six corresponding constraints ($M$, $R$, and $T_\text{eff}$ of both stars), and leaving the initial masses of the two stars and the age of the binary system as the only three free parameters. For both stars, we adopted an overshooting parameter $\alpha_\text{ov} = 0.20$, and no turbulent diffusion. The parameters of the best-fit models are reported in Table\,\ref{table:cles}. Except for the secondary effective temperature, all stellar parameters are well-reproduced within their error bars. The age of the binary system is estimated at $6.25\pm 0.30$\,Myr while the total (Newtonian + general relativistic) apsidal motion rate is estimated at $16\fdg08\,\text{yr}^{-1}$, slightly larger than the observational value (see Sect.\,\ref{sect:times_minima}). We then computed two models independently for each star (Models $\text{II}_\text{P}$ and $\text{II}_\text{S}$) adopting the same constraints and free parameters. The best-fit models in this case (see Table\,\ref{table:cles}) have ages of $6.13\pm0.37$ and $7.47\pm1.30$\,Myr. The physical parameters of the two stars are better reproduced but the apsidal motion rate, of $15\fdg98\,\text{yr}^{-1}$, is still slightly larger than the observational value. We note that in principle, we should compute the apsidal motion rate of the system based on two stellar models having the same age. Nonetheless, the independent primary and secondary models have ages compatible within their error bars. \\

\begin{table*}[h!tb]
\caption{Parameters of some best-fit {\tt Cl\'es} models of CPD-41$^\circ$\,7742 discussed in Sect.\,\ref{sect:cles}.}
\label{table:cles}
\centering
\begin{tabular}{l c c c c c c c c c c}
\hline\hline
\vspace*{-0.3cm} \\
Model &  Age & $M_{\rm init}$ & $M$ & $R$ & $T_{\rm eff}$ &    $k_{2,\text{un.}}$ & $k_2$    & $10^{-8}$\, $\dot{M}$  & $\alpha_\text{ov}$ & $10^6$\,$D_T$\\  
& (Myr) & ($M_{\odot}$) & ($M_{\odot}$)  & ($R_{\odot}$)  & (K)  & ($10^{-3}$)& ($10^{-3}$)  & ($M_{\odot}$\,yr$^{-1}$)  && (cm$^2$\,s$^{-1}$)  \\
\vspace*{-0.3cm} \\
\hline
\vspace*{-0.3cm} \\
Model I$_\text{P}$ & $6.25$ & $17.89$ & $17.73$ & $7.55$ & $31\,150$ &  6.6573 &  6.2738 & 3.67  &0.20&  0 \\
~~~~~~~~~~~\,I$_\text{S}$ & & $10.11$ & $10.11$ & $4.26$ & $25\,292$ &  9.7592 & 9.5318  & $4.10\times 10^{-2}$& 0.20&   0 \\
Model II$_\text{P}$ & $6.13$ & $18.07$ & $17.91$ & $7.57$ & $31\,300$ &  6.7045 & 6.3196 & 3.88  &0.20&  0 \\
Model II$_\text{S}$ & $7.47$ & $9.89$ & $9.89$ & $4.29$ & $24\,926$ &  9.3022 & 9.0757 & $3.54\times 10^{-2}$  &0.20&  0 \\
Model III$_\text{P}$ (1) & $6.83$ & $17.99$ & $17.79$ & $7.57$ & $31\,800$ &  6.0474 & 5.6974  & 4.74  &0.20&  1.81 \\
Model III$_\text{P}$ (2)& $6.81$ & $17.99$ & $17.79$ & $7.57$ & $31\,800$ &  6.0427 & 5.6930 & 4.73  &0.25&  1.41 \\
Model III$_\text{P}$ (3)& $6.80$ & $17.99$ & $17.79$ & $7.57$ & $31\,800$ &  6.0389 & 5.6894  & 4.75  &0.30&  1.05 \\
Model III$_\text{P}$ (4)& $6.79$ & $17.99$ & $17.79$ & $7.57$ & $31\,800$ &  6.0358 &  5.6865 & 4.74  &0.35&  0.74 \\
Model III$_\text{P}$ (5)& $6.78$ & $17.99$ & $17.79$ & $7.57$ & $31\,800$ &  6.0338 &  5.6846 & 4.75  &0.40&  0.48 \\
Model III$_\text{S}$ (1) & $6.99$ & $9.96$ & $9.95$ & $4.27$ & $25\,045$ &  9.4745 & 9.2482  & $3.69\times 10^{-2}$  &0.20&  0.0 \\
Model III$_\text{S}$ (2) & $7.22$ & $9.95$ & $9.94$ & $4.28$ & $25\,060$ &  9.4014 & 9.1759  & $3.74\times 10^{-2}$  &0.25&  0.0 \\
Model III$_\text{S}$ (3)& $7.50$ & $9.92$ & $9.92$ & $4.28$ & $25\,054$ &  9.3132 &  9.0888 & $3.74\times 10^{-2}$  &0.30&  0.0 \\
Model III$_\text{S}$ (4)& $7.72$ & $9.92$ & $9.92$ & $4.29$ & $25\,075$ &  9.2483 & 9.0246  & $3.80\times 10^{-2}$  &0.35&  0.0 \\
Model III$_\text{S}$ (5)& $7.88$ & $9.92$ & $9.91$ & $4.29$ & $25\,104$ &  9.2097 &  8.9869 & $3.87\times 10^{-2}$  &0.40&  0.0 \\
\vspace*{-0.3cm} \\
\hline
\end{tabular}
\tablefoot{Columns\,1 and 2 give the name of the model and its current age. Column\,3 lists the initial mass of the corresponding evolutionary sequence. Columns\,4, 5, and 6 give the mass, radius, and effective temperature. Columns\,7 and 8 yield the $k_2$ respectively before and after applying the empirical correction for the effect of rotation of \citet{Claret99}. Column\,9 lists the mass-loss rate. Columns\,10 and 11 give the overshooting parameter and turbulent diffusion. Models with subscript $\text{P}$ (respectively $\text{S}$) correspond to the models of the primary (respectively secondary) star. 
Models I and II have $D_T$ fixed to 0\,cm$^2$\,s$^{-1}$ while Models III let it vary freely.}
\end{table*}

In a last attempt to reproduce the stellar properties and the apsidal motion rate of the system, we built a series of five independent models (Models III (1) to (5)) for the two stars adopting the same constraints as before, an overshooting parameter of 0.20, 0.25, 0.30, 0.35, and 0.40, respectively, and, as free parameters, the age, the initial mass, and the turbulent diffusion. The best-fit models are reported in Table\,\ref{table:cles}. The models perfectly reproduce the stellar properties of the primary star thanks to the addition of turbulent diffusion. The best-fit turbulent diffusion coefficient decreases with increasing overshooting, indicating that both effects affect the $k_2$ parameter in a similar way \citep[see also the discussion in][]{rosu20b}. 
We note that the turbulent diffusion of the secondary models converge towards a value of 0\,cm$^2$\,s$^{-1}$. This behaviour has already been observed for the binary system HD\,152219 and we refer to the discussion in \citet{rosu22}. The age estimates range from 6.78 to 6.83\,Myr for the primary star and from 6.99 to 7.88\,Myr for the secondary star. For each series of models, we computed the apsidal motion rate of the binary system and obtained values ranging between $14\fdg79$ and $14\fdg89\,\text{yr}^{-1}$. These values are compatible within the error bars with the observational value. We therefore conclude that to reproduce the apsidal motion rate of the binary system, and hence, the internal structure of the stars composing the system, turbulent diffusion needs to be included inside the models, at least for the most massive star of the system. These results confirm those obtained by \citet{rosu20b} for the twin system HD\,152248, and \citet{rosu22} for the primary star of HD\,152219, two systems located in the same cluster NGC\,6231. 

\section{Revisiting HD\,152218\label{sect:HD152218}}
In this section, we briefly reconsider the eccentric massive binary HD\,152218, also located in NGC\,6231. Since our previous work on this system \citep{rauw16}, TESS photometry has become available and the \texttt{min-Cl\'es} routine allows for a more efficient search for best-fit stellar evolution models, which now also includes turbulent diffusion. The stellar and orbital parameters of HD\,152218 \citep[taken from the RV analysis of][]{rauw16} are summarised in Table\,\ref{table:HD152218}. We determined a Newtonian and a general relativistic contributions to the apsidal motion rate of $1\fdg91^{+0.23}_{-0.24}$\,yr$^{-1}$ and $0\fdg13\pm0\fdg01\,\text{yr}^{-1}$, respectively. 

\begin{table}[h]
\caption{Observational properties of the binary system HD\,152218 taken from \citet{rauw16}.\label{table:HD152218}}
\centering
\begin{tabular}{l l l}
\hline\hline
\vspace{-3mm}\\
Parameter & \multicolumn{2}{c}{Value} \\
& Primary & Secondary \\
\hline
\vspace{-3mm}\\
$M$ ($M_\odot$) & $19.8 \pm 1.5$ & $15.0 \pm 1.1$  \\
\vspace{-3mm}\\
$R$ ($R_\odot$) & $8.4^{+1.1}_{-1.0}$ & $7.8^{+1.0}_{-1.6}$ \\
\vspace{-3mm}\\
$T_\text{eff}$ (K) & $33\,400 \pm 1000$ & $29\,900\pm 1000$ \\
\vspace{-3mm}\\
$P_\text{rot}$ (d) & $2.69^{+0.37}_{-0.34}$ & $2.56^{+0.48}_{-0.63}$ \\
\vspace{-3mm}\\
$e$ & \multicolumn{2}{c}{$0.280^{+0.010}_{-0.008}$} \\
\vspace{-3mm}\\
$i$  & \multicolumn{2}{c}{$66\fdg3^{+3.0}_{-3.3}$} \\
\vspace{-3mm}\\
$P_\text{orb}$ (d)& \multicolumn{2}{c}{$5.60445 \pm 0.00005$} \\
\vspace{-3mm}\\
$\dot\omega$ ($^\circ\,\text{yr}^{-1}$)& \multicolumn{2}{c}{$2.04^{+0.23}_{-0.24}$} \\
\vspace{-3mm}\\
\hline
\end{tabular}
\tablefoot{We note there was a small typo in the secondary rotational period in \citet{rauw16}, which we corrected here.}
\end{table} 

\subsection{Analysis of TESS data}
HD\,152218 was observed by TESS during the same sectors as CPD-41$^\circ$\,7742 (i.e. sectors 12 and 39).  We performed an extraction of the light curves with a 30 and 10 minute cadences again for TESS-12 and TESS-39, respectively. For this purpose, we used the {\tt Lightkurve} software and followed the reduction steps explained in Sect.\,\ref{subsect:obs_photo}. We adopted here the pca-5 for the background subtraction. This results in a total of 1257 and 3705 data points for TESS-12 and TESS-39, respectively.

\begin{figure}[h!]
\includegraphics[clip=true, trim=35 40 40 200,width=\linewidth]{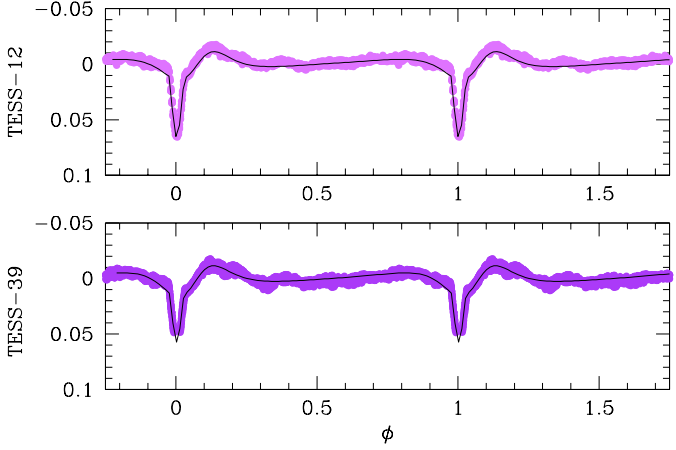}
\caption{Best-fit solution to the TESS-12 and TESS-39 photometry of HD\,152218. The light curves have been phase-folded with $P_\text{ecl} = 5.60415$\,d computed using Eq.\,\eqref{eqn:Pecl}. \label{fig:nightfall_HD152218}}
\end{figure}

We analysed the light curves of HD\,152218 with the \texttt{Nightfall} code. We fixed the effective temperatures, the eccentricity, and the orbital period to the values quoted by \citet[][see Table\,\ref{table:HD152218}]{rauw16}. The only free parameters were the orbital inclination, the longitude of periastron, and the Roche lobe filling factors of the two stars. The best-fit adjustments are given in Table\,\ref{table:nightfall_HD152218} (we note again that the large values of $\chi_\nu^2$ have no physical sense and arise because of the underestimate of the errors on the TESS data) and illustrated in Fig.\,\ref{fig:nightfall_HD152218}. 

\begin{table}[h]
\centering
\caption{Best-fit parameters of the TESS photometry of HD\,152218. \label{table:nightfall_HD152218}}
\begin{tabular}{l l l l l}
\hline\hline
\vspace*{-3mm}\\
Parameter & \multicolumn{2}{c}{Value} \\
& TESS-12 & TESS-39  \\
\hline
\vspace*{-3mm}\\
Epoch (HJD) & 2\,458\,638.98 & 2\,459\,375.68 \\
\vspace*{-3mm}\\
$i$& $68\fdg06^{+0.6}_{-0.40}$ & $66\fdg09^{+1.60}_{-0.40}$ \\
\vspace*{-3mm}\\
$f_\text{P}$ & $0.673^{+0.020}_{-0.022}$  &$0.680^{+0.035}_{-0.028}$ \\
\vspace*{-3mm}\\
$f_\text{S}$ & $0.679_{-0.023}^{+0.026}$  &$0.721^{+0.028}_{-0.036}$  \\
\vspace*{-3mm}\\
$\omega$  & $132\fdg2^{+0.6}_{-1.1}$ & $124\fdg9^{+0.8}_{-1.2}$ \\
\vspace*{-3mm}\\
$\chi^2_\nu$ & 1042.8 & 437.8 \\
\vspace*{-3mm}\\
\hline
\vspace*{-3mm}\\
$\omega$-range$^a$ & $125\fdg5$ -- $133\fdg8$ & $129\fdg2$ -- $137\fdg9$ \\
\vspace*{-3mm}\\
\hline
\end{tabular}
\tablefoot{$^a$Range of $\omega$-values that give acceptable results in terms of $\chi^2_\nu$ when $i$, $f_\text{P}$, and $f_\text{S}$ are fixed to the values that, considering Fig.\,6 of \citet{rauw16}, give a brightness ratio in accordance with the spectroscopic value and are acceptable (at $1\sigma$) with the ASAS-3 photometry.}
\end{table}

The values for the longitude of periastron expected at the times of the TESS-12 and TESS-39 observations from the apsidal motion rate derived from the RV analysis of \citet{rauw16} amount to $135\fdg8^{+5.4}_{-5.7}$ and $139\fdg9^{+5.9}_{-6.1}$, respectively. Whilst the value of $\omega$ inferred for the TESS-12 light curve is compatible within the error bars, this is not the case for the TESS-39 light curve. In addition, the value of $\omega$ should be larger for the TESS-39 light curve than for the TESS-12 one. 

In the case of HD\,152218, the fact that there is a single eclipse leads to a degeneracy between the primary and secondary Roche lobe filling factors, and the orbital inclination that can shift the formally best-fit solution to values of the Roche lobe filling factors that are inconsistent with the spectroscopic brightness ratio. To overcome this difficulty, we scanned the parameter space fixing $i, f_\text{P}$, and $f_\text{S}$ to the values that, considering Fig.\,6 of \citet{rauw16}, give a brightness ratio in accordance with the spectroscopic value and are acceptable (at $1\sigma$) with the ASAS-3 photometry. In this way, $\omega$ was the only free parameter of the adjustment. The combination of parameters that give acceptable results in terms of $\chi_\nu^2$ give $\omega$ values ranging from $125\fdg5$ to $133\fdg8$ for the TESS-12 data and from $129\fdg2$ to $137\fdg9$ for the TESS-39 data. These values are in agreement, within the error bars, with those expected from the spectroscopic apsidal motion rate. These solutions all predict an inclination between $67^\circ$ and $69^\circ$.

We therefore conclude that the previously reported apsidal motion rate derived by \citet{rauw16} is confirmed by the new photometric data. 

\subsection{Stellar evolution models\label{subsect:HD152218_cles}}
We built stellar evolution models with the \texttt{min-Cl\'es} routine for HD\,152218, adopting as constraints the masses, radii, and effective temperatures of the stars. We fixed $\xi=1$, and left the age, initial masses, and turbulent diffusion as free parameters.  \\

We first build two models (Models I$_\text{P}$ and I$_\text{S}$) simultaneously for the two stars. The parameters of the best fit are reported in Table\,\ref{table:cles_HD152218}. Except for the radius of the secondary star, all stellar parameters are well-reproduced within their error bars. The age of the binary system is estimated at 5.79\,Myr, confirming the value obtained by \citet{rauw16}. The apsidal motion rate is estimated at $1\fdg60\,\text{yr}^{-1}$. This value is slightly lower than the observational value. \\

\begin{table*}[h!tb]
\caption{Parameters of some best-fit {\tt Cl\'es} models of HD\,152218 discussed in Sect.\,\ref{subsect:HD152218_cles}.}
\label{table:cles_HD152218}
\centering
\begin{tabular}{l c c c c c c c c c c}
\hline\hline
\vspace*{-0.3cm} \\
Model &  Age & $M_{\rm init}$ & $M$ & $R$ & $T_{\rm eff}$ &    $k_{2,\text{un.}}$ & $k_2$    & $10^{-8}$\, $\dot{M}$  & $\alpha_\text{ov}$ & $10^6$\,$D_T$  \\
& (Myr) & ($M_{\odot}$) & ($M_{\odot}$)  & ($R_{\odot}$)  & (K)  & ($10^{-3}$)& ($10^{-3}$)  & ($M_{\odot}$\,yr$^{-1}$)  && (cm$^2$\,s$^{-1}$)  \\
\vspace*{-0.3cm} \\
\hline
\vspace*{-0.3cm} \\
Model I$_\text{P}$ & $5.79$ & $20.02$ & $19.72$ & $8.12$ & $32\,696$ &  6.0908 &  5.6951 & 7.81  &0.20&  1.0 \\
~~~~~~~~~~~\,I$_\text{S}$& & $15.30$ & $15.25$ & $6.08$ & $30\,134$ &  8.6278 & 8.2863  & 1.15& 0.20&   0.0 \\
Model II$_\text{P}$ (1) & $6.76$ & $20.23$ & $19.80$ & $8.40$ & $33\,400$ &  4.9192 & 4.5680  & 11.5  &0.20&  2.07 \\
Model II$_\text{P}$ (2)& $6.75$ & $20.23$ & $19.80$ & $8.40$ & $33\,400$ &  4.9106 &  4.5600 & 11.5  &0.25&  1.65 \\
Model II$_\text{P}$ (3)& $6.73$ & $20.23$ & $19.80$ & $8.40$ & $33\,400$ &  4.9027 &  4.5527 & 11.5  &0.30&  1.27 \\
Model II$_\text{P}$ (4)& $6.72$ & $20.23$ & $19.80$ & $8.40$ & $33\,400$ &  4.8957 &  4.5462 & 11.5  &0.35&  0.94 \\
Model II$_\text{P}$ (5)& $6.71$ & $20.23$ & $19.80$ & $8.40$ & $33\,400$ &  4.8892 &  4.5402 & 11.5  &0.40&  0.65 \\
Model II$_\text{S}$ (1) & $11.67$ & $15.17$ & $15.00$ & $7.80$ & $29\,900$ &  3.9358 & 3.6090  & 3.19  &0.20&  4.15 \\
Model II$_\text{S}$ (2) & $11.65$ & $15.17$ & $15.00$ & $7.80$ & $29\,900$ &  3.9264 & 3.6004  & 3.19  &0.25&  3.75 \\
Model II$_\text{S}$ (3)& $11.62$ & $15.17$ & $15.00$ & $7.80$ & $29\,900$ &  3.9179 &  3.5926 & 3.20  &0.30&  3.37 \\
Model II$_\text{S}$ (4)& $11.60$ & $15.17$ & $15.00$ & $7.80$ & $29\,900$ &  3.9100 &  3.5853 & 3.19  &0.35&  3.01 \\
Model II$_\text{S}$ (5)& $11.59$ & $15.17$ & $15.00$ & $7.80$ & $29\,900$ &  3.9028 &  3.5787 & 3.19  &0.40&  2.67 \\
Model III$_\text{S}$ (1) & $9.92$ & $15.11$ & $15.00$ & $7.12$ & $29\,900$ &  5.2520 &  4.9169 & 2.13  &0.20&  2.51 \\
Model III$_\text{S}$ (2) & $9.90$ & $15.11$ & $15.00$ & $7.12$ & $29\,900$ &  5.2450 &  4.9103 & 2.14  &0.25&  2.17 \\
Model III$_\text{S}$ (3)& $9.88$ & $15.11$ & $15.00$ & $7.12$ & $29\,900$ &  5.2387 & 4.9044  & 2.14  &0.30&  1.84 \\
Model III$_\text{S}$ (4)& $9.86$ & $15.11$ & $15.00$ & $7.12$ & $29\,900$ &  5.2331 &  4.8992 & 2.13  &0.35&  1.54 \\
Model III$_\text{S}$ (5)& $9.85$ & $15.11$ & $15.00$ & $7.12$ & $29\,900$ &  5.2281 & 4.8945  & 2.14  &0.40&  1.26 \\
Model IV$_\text{P}$ (1) & $4.98$ & $20.69$ & $20.41$ & $7.98$ & $33\,071$ &  6.9772 & 6.5606  & 7.90  &0.20&  0 \\
Model IV$_\text{P}$ (2)& $5.09$ & $20.66$ & $20.37$ & $8.01$ & $33\,090$ &  6.8012 &  6.3895 & 8.12  &0.25&  0 \\
Model IV$_\text{P}$ (3)& $5.20$ & $20.63$ & $20.33$ & $8.04$ & $33\,110$ &  6.6310 &  6.2241 & 8.34  &0.30&  0 \\
Model IV$_\text{P}$ (4)& $5.31$ & $20.60$ & $20.29$ & $8.08$ & $33\,130$ &  6.4662 &  6.0641 & 8.56  &0.35&  0 \\
Model IV$_\text{P}$ (5)& $5.42$ & $20.57$ & $20.26$ & $8.11$ & $33\,151$ &  6.3054 &  5.9082 & 8.79  &0.40&  0 \\
Model IV$_\text{S}$ (1) & $7.39$ & $15.53$ & $15.45$ & $6.87$ & $29\,554$ &  6.8463 & 6.4640  & 1.52  &0.20& 0 \\
Model IV$_\text{S}$ (2) & $7.61$ & $15.52$ & $15.43$ & $6.92$ & $29\,566$ &  6.6354 & 6.2556  & 1.59  &0.25& 0\\
Model IV$_\text{S}$ (3)& $7.84$ & $15.50$ & $15.42$ & $6.98$ & $29\,580$ &  6.4322 &  6.0550 & 1.65  &0.30& 0 \\
Model IV$_\text{S}$ (4)& $8.06$ & $15.49$ & $15.40$ & $7.03$ & $29\,594$ &  6.2343 &  5.8599 & 1.72  &0.35& 0 \\
Model IV$_\text{S}$ (5)& $8.29$ & $15.47$ & $15.37$ & $7.09$ & $29\,608$ &  6.0421 &  5.6706 & 1.80  &0.40& 0 \\
Model V$_\text{S}$ (1) & $6.98$ & $15.35$ & $15.28$ & $6.56$ & $29\,687$ &  7.3970 &  7.0319 & 1.34  &0.20& 0 \\
Model V$_\text{S}$ (2) & $7.20$ & $15.33$ & $15.26$ & $6.60$ & $29\,700$ &  7.1912 &  6.8282 & 1.39  &0.25& 0\\
Model V$_\text{S}$ (3)& $7.42$ & $15.32$ & $15.24$ & $6.66$ & $29\,714$ &  6.9904 & 6.6295  & 1.44  &0.30&  0 \\
Model V$_\text{S}$ (4)& $7.65$ & $15.30$ & $15.22$ & $6.70$ & $29\,728$ &  6.7955 &  6.4370 & 1.50  &0.35&  0 \\
Model V$_\text{S}$ (5)& $7.88$ & $15.28$ & $15.20$ & $6.75$ & $29\,744$ &  6.6039 & 6.2478  & 1.56  &0.40&  0 \\
\vspace*{-0.3cm} \\
\hline
\end{tabular}
\tablefoot{Columns\,1 and 2 give the name of the model and its current age. Column\,3 lists the initial mass of the corresponding evolutionary sequence. Columns\,4, 5, and 6 give the mass, radius, and effective temperature. Columns\,7 and 8 yield the $k_2$ respectively before and after applying the empirical correction for the effect of rotation of \citet{Claret99}. Column\,9 lists the mass-loss rate. Columns\,10 and 11 give the overshooting parameter and turbulent diffusion. Models with subscript $\text{P}$ (respectively $\text{S}$) correspond to the models of the primary (respectively secondary) star. 
Models IV and V have $D_T$ fixed to 0\,cm$^2$\,s$^{-1}$ while Models I, II, and III allow it to vary freely.}
\end{table*}

To solve the discrepancies in the secondary radius and in the apsidal motion rate, we further built a series of five independent models (Models $\text{II}_\text{P}$ (1) to (5) and $\text{II}_\text{S}$ (1) to (5)) for the two stars adopting an overshooting parameter of 0.20, 0.25, 0.30, 0.35, and 0.40, respectively. The best-fit models are reported in Table\,\ref{table:cles_HD152218}. They reproduce well the stellar properties of both stars. As for the primary star of CPD-41$^\circ$\,7742, we observe that the amount of turbulent diffusion necessary to fit the stellar properties decreases with the increasing overshooting parameter. The age estimates range from 6.71 to 6.76\,Myr for the primary star and from 11.59 to 11.67\,Myr for the secondary star. Given the important age difference between the primary and secondary models, we cannot compute a coherent apsidal motion rate for the binary system by simply summing the contributions of the two stars taken at different ages. \\

As discussed in \citet{rauw16}, the presence of only one eclipse in the light curves of HD\,152218 leads to degeneracies in the determination of the stellar radii, the ratio of effective temperatures, and the inclination of the system. \citet{rauw16} derived a ratio of $1.18 \pm 0.20$ between the primary and secondary radii based on their ratio of bolometric luminosities and effective temperatures. Assuming that the radius of the primary star is well constrained, this relation translates into a secondary radius of $7.12\,R_\odot$. Thus, we built a series of five models for the secondary star assuming a constraint on the stellar radius of 7.12 rather than $7.80\,R_\odot$. These best-fit models (Models III$_\text{S}$ (1) to (5)), which reproduce well the stellar properties, are reported in Table\,\ref{table:cles_HD152218}. Compared to Models II$_\text{S}$, the age estimate is lower and closer to that of the primary star models, and the apsidal motion rate, computed based on Models II$_\text{P}$ and III$_\text{S}$, amounts to $1\fdg62$\,yr$^{-1}$ to $1\fdg63$\,yr$^{-1}$.  \\

We then built a \texttt{Cl\'es} evolutionary sequence for the secondary star, starting with an initial mass of $15.17\,M_\odot$ and having $\alpha_\text{ov}=0.20$ and no turbulent diffusion. We stopped the sequence at the age of the best-fit models of the primary star (Models II$_\text{P}$) and obtained a model of mass $15.11\,M_\odot$, radius $6.39\,R_\odot$, effective temperature 29\,674\,K, and $k_{2,2}=7.68\times 10^{-3}$. Combined with Model II$_\text{P}$ (1) for the primary star, we obtained an apsidal motion rate of $1\fdg60\,\text{yr}^{-1}$. This value is slightly lower than the observational value but agrees with the previously obtained value. \\

Finally, in a last attempt to reproduce the apsidal motion rate of the system, we built models for the two stars assuming $D_T=0$\,cm$^2$\,s$^{-1}$ (Models IV$_\text{P}$ for the primary star, Models IV$_\text{S}$ and V$_\text{S}$ for the secondary star, when the constraint on the radius is 7.80 and $7.12\,R_\odot$, respectively). The stellar properties are still well reproduced within the error bars, and the apsidal motion rate (neglecting the age difference between the primary and secondary models) amounts to $1\fdg81$\,yr$^{-1}$ to $1\fdg84$\,yr$^{-1}$ and $1\fdg72$\,yr$^{-1}$ to $1\fdg75$\,yr$^{-1}$ when Models IV$_\text{S}$ and V$_\text{S}$ are considered. This value is compatible within the error bars with the observational value. Initial masses of $20.6\pm1.5$ and $15.5\pm 1.1$\,$M_\odot$ are derived for the primary and secondary stars, respectively, and the binary's age is estimated at $5.2\pm 0.8$\,Myr as derived from the primary models. 

\section{Conclusion \label{sect:conclusion}}
The eccentric massive binary CPD-41$^\circ$\,7742, belonging to the open cluster NGC\,6231, has been analysed both from an observational and a stellar evolution point of views. The spectroscopic observations were used to obtain the reconstructed spectra of the components together with their RVs at each time of observation. Combined with the RVs coming from the literature, the latter were subsequently used to determine stellar and orbital parameters for the system.  We further analysed the photometric observations of the system to constrain the inclination of the orbit as well as the absolute masses and radii of the stars. We solved the inconsistency highlighted by \citet{sana05} and \citet{bouzid05} between the longitude of periastron inferred by the RV analysis on the one hand and the photometric analysis on the other hand by analysing the phase differences between the times of secondary and primary minimum of the eclipses for the different sets of photometric observations explicitly accounting for the apsidal motion. In this way, we obtained a rate of apsidal motion of $(15\fdg38^{+0.42}_{-0.51})\,\text{yr}^{-1}$. We then computed stellar evolution models for the two stars. The best-fit models of the stars, in terms of the masses, radii, and effective temperatures, predict a theoretical apsidal motion rate compatible within the error bars with the observed one provided that turbulent mixing is introduced in the primary stellar evolution models, therefore confirming the results obtained for the twin system HD\,152248 by \citet{rosu20b} and for HD\,152219 by \citet{rosu22}. The binary's age is estimated at $6.8 \pm 1.4$\,Myr and initial masses of $18.0 \pm 0.5$ and $9.9 \pm 0.3$\,$M_\odot$ are derived for the primary and secondary stars, respectively.  \\

The eccentric massive binary HD\,152218, belonging to the same cluster, was reconsidered in light of the newly available TESS data and the updated \texttt{Cl\'es} code. The results of the light curve analysis are coherent with those obtained from our previous study \citep{rauw16}.  
The stellar evolution models reproduce the observational properties of the stars within their error bars, including the apsidal motion rate, and predict initial masses of $20.6\pm1.5$ and $15.5\pm 1.1$\,$M_\odot$ for the primary and secondary stars, respectively, as well as a binary age of $5.2\pm 0.8$\,Myr.\\

In total, we have analysed four massive eccentric binary systems belonging to the NGC\,6231 cluster in a series of papers \citep[see][]{Rosu, rosu20b, rosu22} and the present work. The ages estimated for these systems point towards an age estimate for the massive star population of the cluster of 5 to 9.5\,Myr. This value is compatible with the ages estimated (between 1 to 7\,Myr with a small peak at 3\,Myr) based on the low-mass pre-main-sequence stars of the cluster, though pointing towards the upper allowed range of values. The large scatter is partially due to the observational uncertainties (e.g. the lack of the secondary eclipse in HD\,152218 induces degeneracies in the determination of the stellar parameters) as well as due to physical assumption lying behind the stellar evolution models. These studies confirm the need for enhanced mixing inside the stellar evolution models to reproduce the observational stellar properties. They suggest that massive stars have larger convective cores than usually considered in stellar evolution codes.

 \begin{acknowledgements}
S. R., Y. N., and E. G. acknowledge support from the Fonds de la Recherche Scientifique (F.R.S.- FNRS, Belgium). We thank Drs John Hillier and Rainer Wichmann for making their codes, respectively {\tt CMFGEN} and {\tt Nightfall}, publicly available. We thank Dr Martin Farnir for the development of the \texttt{min-Cl\'es} routine and his help regarding its use. The authors thank the referee for his/her suggestions and comments towards the improvement of the manuscript.
\end{acknowledgements}

\onecolumn
\begin{appendix} 
\section{Journal of the spectroscopic observations of CPD-41$^\circ$\,7742 \label{appendix:spectrotable}}
This appendix provides the journal of the spectroscopic observations of CPD-41$^\circ$\,7742 (Table\,\ref{Table:spectro+RV}).

\samepage
\longtab[1]{
\begin{longtable}{ccrrl}
\caption{Journal of the spectroscopic observations of CPD-41$^\circ$\,7742. \label{Table:spectro+RV}}\\
\hline
\vspace{-3mm}\\
HJD & $\phi$  & $RV_\text{P}$ & $RV_\text{S}$ & Inst.\\
--\,2\,450\,000&  & (km\,s$^{-1}$) & (km\,s$^{-1}$) & \\
\hline
\vspace{-3mm}\\
\endfirsthead 
\caption{continued.}\\
\hline
\hline
\vspace{-3mm}\\
HJD & $\phi$  & $RV_\text{P}$ & $RV_\text{S}$ & Inst.\\
--\,2\,450\,000&  & (km\,s$^{-1}$) & (km\,s$^{-1}$) & \\
\hline
\vspace{-3mm}\\
\endhead 
\hline
\endfoot
\hline
\endlastfoot
1299.843 & 0.172 & $ -183.3 \pm 1.2 $ & $  264.2 \pm 5.7 $ & F\\
1300.841 & 0.581 & $  138.3 \pm 1.4 $ & $ -303.7 \pm 4.1 $ & F\\
1301.847 & 0.993 & $ -129.9 \pm 2.3 $ & $  200.5 \pm 8.9 $ & F\\
1302.847 & 0.403 & $    8.5 \pm 2.5 $ & $ -102.0 \pm 28.1 $ & F\\
1304.841 & 0.219 & $ -154.4 \pm 2.5 $ & $  261.1 \pm 12.0 $ & F\\
1327.863 & 0.652 & $  146.7 \pm 1.4 $ & $ -325.2 \pm 5.6 $ & F\\
1668.928 & 0.387 & $    0.5 \pm 1.8 $ & $  -62.7 \pm 9.9 $ & F\\
1669.791 & 0.741 & $  108.2 \pm 1.0 $ & $ -251.6 \pm 3.0 $ & F\\
1669.923 & 0.795 & $   62.8 \pm 1.5 $ & $ -172.3 \pm 6.2 $ & F\\
1670.787 & 0.149 & $ -187.6 \pm 1.3 $ & $  263.7 \pm 5.6 $ & F\\
1670.901 & 0.196 & $ -171.8 \pm 1.5 $ & $  241.5 \pm 4.7 $ & F\\
1671.800 & 0.564 & $  133.5 \pm 1.1 $ & $ -298.0 \pm 4.7 $ & F\\
1671.926 & 0.615 & $  147.8 \pm 1.1 $ & $ -318.4 \pm 5.2 $ & F\\
1672.782 & 0.966 & $ -117.6 \pm 1.2 $ & $  147.4 \pm 2.2 $ & F\\
1672.925 & 0.025 & $ -161.3 \pm 1.5 $ & $  222.5 \pm 3.3 $ & F\\
2037.792 & 0.512 & $  110.4 \pm 1.1 $ & $ -259.8 \pm 3.6 $ & F\\
2037.887 & 0.551 & $  128.4 \pm 1.1 $ & $ -283.8 \pm 4.0 $ & F\\
2039.783 & 0.328 & $  -65.6 \pm 0.8 $ & $   13.2 \pm 7.8 $ & F\\
2039.907 & 0.379 & $  -12.4 \pm 1.0 $ & $   -8.1 \pm 7.3 $ & F\\
2040.788 & 0.739 & $  103.2 \pm 1.4 $ & $ -238.7 \pm 2.1 $ & F\\
2040.908 & 0.789 & $   61.2 \pm 2.1 $ & $ -161.7 \pm 6.8 $ &F\\
2337.860 & 0.451 & $   70.6 \pm 1.7 $ & $ -189.0 \pm 6.8 $ & F\\
2338.794 & 0.834 & $   23.5 \pm 0.5 $ & $  -77.7 \pm 9.2 $ & F\\
2381.723 & 0.422 & $   48.1 \pm 1.3 $ & $ -135.0 \pm 5.0 $ & F\\
2382.741 & 0.839 & $   16.9 \pm 0.8 $ & $  -71.9 \pm 7.9 $ & F\\
2383.738 & 0.247 & $ -126.1 \pm 1.1 $ & $  172.4 \pm 2.7 $ & F\\
4924.888 & 0.365 & $   66.0 \pm 1.5 $ & $ -160.8 \pm 4.4 $ & H\\
4924.909 & 0.374 & $   73.6 \pm 1.5 $ & $ -177.8 \pm 3.4 $ & H\\
4925.655 & 0.679 & $   84.4 \pm 1.4 $ & $ -206.1 \pm 2.0 $ & H\\
4925.927 & 0.791 & $  -40.9 \pm 2.8 $ & $   53.9 \pm 5.6 $ &H\\
4926.708 & 0.111 & $ -174.9 \pm 1.1 $ & $  255.0 \pm 3.1 $ & H\\
4926.712 & 0.112 & $ -174.2 \pm 1.0 $ & $  250.0 \pm 1.6 $ & H\\
4926.758 & 0.131 & $ -163.0 \pm 1.0 $ & $  237.1 \pm 1.6 $ & H\\
4926.762 & 0.133 & $ -162.1 \pm 1.3 $ & $  233.9 \pm 1.7 $ & H\\
4926.820 & 0.157 & $ -146.0 \pm 1.0 $ & $  205.5 \pm 1.7 $ & H\\
4926.840 & 0.165 & $ -139.9 \pm 1.2 $ & $  198.0 \pm 2.7 $ & H\\
4926.844 & 0.166 & $ -139.0 \pm 1.0 $ & $  192.8 \pm 1.3 $ & H\\
4926.913 & 0.195 & $ -113.9 \pm 1.2 $ & $  149.0 \pm 3.7 $ & H\\
4927.661 & 0.501 & $  142.9 \pm 1.2 $ & $ -310.6 \pm 4.8 $ & H\\
4927.665 & 0.503 & $  142.5 \pm 1.2 $ & $ -314.1 \pm 3.6 $ & H\\
4927.737 & 0.532 & $  146.2 \pm 1.0 $ & $ -313.7 \pm 3.1 $ & H\\
4927.741 & 0.534 & $  145.1 \pm 1.1 $ & $ -318.4 \pm 3.6 $ & H\\
4927.793 & 0.555 & $  143.1 \pm 1.2 $ & $ -316.4 \pm 1.8 $ & H\\
4927.797 & 0.557 & $  144.3 \pm 1.2 $ & $ -311.2 \pm 3.4 $ & H\\
4927.847 & 0.577 & $  140.0 \pm 1.3 $ & $ -302.6 \pm 2.8 $ & H\\
4927.851 & 0.579 & $  139.0 \pm 1.5 $ & $ -301.8 \pm 4.6 $ & H\\
4927.855 & 0.581 & $  138.0 \pm 1.1 $ & $ -297.7 \pm 6.4 $ & H\\
4927.902 & 0.600 & $  131.9 \pm 1.5 $ & $ -285.5 \pm 8.2 $ & H\\
4927.905 & 0.601 & $  131.1 \pm 1.1 $ & $ -286.8 \pm 2.4 $ & H\\
4928.672 & 0.915 & $ -144.9 \pm 1.4 $ & $  199.3 \pm 3.9 $ & H\\
4928.676 & 0.917 & $ -145.4 \pm 1.6 $ & $  199.8 \pm 2.0 $ & H\\
4928.754 & 0.949 & $ -166.7 \pm 1.5 $ & $  237.1 \pm 3.1 $ & H\\
4928.758 & 0.951 & $ -167.3 \pm 1.2 $ & $  243.7 \pm 4.4 $ & H\\
4928.837 & 0.983 & $ -184.4 \pm 1.2 $ & $  262.9 \pm 4.1 $ & H\\
4928.838 & 0.983 & $ -184.4 \pm 1.1 $ & $  269.0 \pm 2.7 $ & H\\
4928.896 & 0.007 & $ -190.8 \pm 1.4 $ & $  273.5 \pm 3.6 $ & H\\
4928.900 & 0.009 & $ -191.3 \pm 1.4 $ & $  275.8 \pm 3.2 $ & H\\
4928.931 & 0.022 & $ -191.7 \pm 1.4 $ & $  283.1 \pm 4.2 $ & H\\
4928.935 & 0.023 & $ -192.5 \pm 1.2 $ & $  281.6 \pm 5.6 $ & H\\
4929.811 & 0.382 & $   79.7 \pm 1.3 $ & $ -189.6 \pm 2.2 $ & H\\
4929.815 & 0.384 & $   81.5 \pm 1.3 $ & $ -190.1 \pm 2.1 $ & H\\
4929.866 & 0.405 & $   97.8 \pm 1.3 $ & $ -229.7 \pm 2.2 $ & H\\
4929.870 & 0.406 & $   99.5 \pm 1.1 $ & $ -222.1 \pm 3.1 $ & H\\
4929.908 & 0.422 & $  110.4 \pm 1.3 $ & $ -245.8 \pm 3.3 $ & H\\
4929.911 & 0.423 & $  111.7 \pm 1.5 $ & $ -243.9 \pm 5.0 $ & H\\
4929.934 & 0.432 & $  116.6 \pm 1.2 $ & $ -261.9 \pm 2.3 $ & H\\
4929.938 & 0.434 & $  116.7 \pm 1.3 $ & $ -263.8 \pm 1.6 $ &H\\
7845.767 & 0.059 & $ -154.7 \pm 1.1 $ & $  213.5 \pm 2.4 $ & G\\
7845.774 & 0.062 & $ -152.7 \pm 1.1 $ & $  210.1 \pm 2.4 $ & G\\
7845.782 & 0.065 & $ -151.2 \pm 1.2 $ & $  206.2 \pm 2.6 $ & G\\
7919.650 & 0.329 & $  104.6 \pm 0.5 $ & $ -241.8 \pm 2.8 $ & G\\
7919.655 & 0.331 & $  106.0 \pm 0.5 $ & $ -245.3 \pm 2.5 $ & G\\
7919.661 & 0.334 & $  107.8 \pm 0.4 $ & $ -247.7 \pm 3.2 $ & G\\
7919.666 & 0.336 & $  108.9 \pm 0.5 $ & $ -250.6 \pm 2.2 $ & G\\
7920.685 & 0.753 & $  -86.2 \pm 0.3 $ & $  101.4 \pm 6.9 $ & G\\
7920.690 & 0.755 & $  -86.8 \pm 0.6 $ & $  110.1 \pm 11.0 $ &G\\
7920.696 & 0.758 & $  -88.2 \pm 0.5 $ & $  113.1 \pm 11.1 $ & G\\
7920.701 & 0.760 & $  -89.6 \pm 0.5 $ & $  112.0 \pm 6.9 $ & G\\
7926.534 & 0.150 & $  -83.1 \pm 1.0 $ & $   83.7 \pm 0.2 $ & G\\
7926.540 & 0.152 & $  -81.6 \pm 1.0 $ & $   80.7 \pm 0.5 $ & G\\
7926.545 & 0.154 & $  -80.2 \pm 0.9 $ & $   77.8 \pm 0.3 $ & G\\
7926.551 & 0.157 & $  -78.6 \pm 1.1 $ & $   72.1 \pm 0.1 $ & G\\
7929.683 & 0.440 & $  151.5 \pm 0.6 $ & $ -332.1 \pm 1.5 $ & G\\
7929.688 & 0.442 & $  151.6 \pm 0.3 $ & $ -331.1 \pm 1.4 $ & G\\
7929.694 & 0.444 & $  151.5 \pm 0.8 $ & $ -332.6 \pm 1.5 $ & G\\
7929.699 & 0.446 & $  151.5 \pm 0.8 $ & $ -330.3 \pm 1.6 $ & G\\
7932.578 & 0.626 & $   56.8 \pm 0.6 $ & $ -153.1 \pm 2.2 $ & G\\
7932.583 & 0.628 & $   54.8 \pm 0.6 $ & $ -151.4 \pm 1.7 $ & G\\
7932.589 & 0.630 & $   53.4 \pm 0.7 $ & $ -146.2 \pm 1.5 $ & G\\
7932.594 & 0.633 & $   51.7 \pm 0.6 $ & $ -142.4 \pm 2.0 $ & G\\
7934.725 & 0.506 & $  140.5 \pm 0.6 $ & $ -309.5 \pm 2.4 $ & G\\
7934.730 & 0.508 & $  139.5 \pm 0.6 $ & $ -307.0 \pm 2.4 $ & G\\
7934.736 & 0.510 & $  138.3 \pm 0.6 $ & $ -305.2 \pm 2.3 $ & G\\
7934.741 & 0.512 & $  137.7 \pm 0.5 $ & $ -303.2 \pm 3.0 $ & G\\
7946.706 & 0.414 & $  147.2 \pm 0.9 $ & $ -326.6 \pm 1.6 $ & G\\
7946.709 & 0.415 & $  147.5 \pm 0.8 $ & $ -326.4 \pm 0.5 $ & G\\
7946.712 & 0.417 & $  148.3 \pm 0.8 $ & $ -327.0 \pm 0.1 $ & G\\
7946.715 & 0.418 & $  148.3 \pm 0.9 $ & $ -327.4 \pm 1.5 $ & G\\
7947.640 & 0.797 & $ -116.8 \pm 0.9 $ & $  157.2 \pm 0.7 $ & G\\
7947.643 & 0.798 & $ -118.1 \pm 1.0 $ & $  159.1 \pm 0.8 $ & G\\
7947.646 & 0.799 & $ -119.4 \pm 0.9 $ & $  160.4 \pm 0.9 $ & G\\
7947.649 & 0.801 & $ -120.2 \pm 1.1 $ & $  163.3 \pm 1.6 $ & G\\
7950.644 & 0.028 & $ -166.5 \pm 1.1 $ & $  243.0 \pm 2.8 $ & G\\
7950.647 & 0.029 & $ -166.2 \pm 1.0 $ & $  241.3 \pm 3.0 $ & G\\
7950.651 & 0.031 & $ -165.7 \pm 1.2 $ & $  241.6 \pm 3.2 $ & G\\
7950.654 & 0.032 & $ -165.1 \pm 1.3 $ & $  240.0 \pm 2.3 $ & G\\
7955.622 & 0.067 & $ -146.1 \pm 0.9 $ & $  202.6 \pm 1.7 $ & G\\
7955.625 & 0.068 & $ -145.1 \pm 1.0 $ & $  201.2 \pm 2.1 $ & G\\
7955.628 & 0.070 & $ -144.4 \pm 1.1 $ & $  201.6 \pm 1.2 $ & G\\
7955.631 & 0.071 & $ -143.2 \pm 1.2 $ & $  199.0 \pm 1.5 $ & G\\
7957.713 & 0.924 & $ -186.1 \pm 1.0 $ & $  265.7 \pm 2.4 $ & G\\
7957.719 & 0.926 & $ -186.1 \pm 1.2 $ & $  266.5 \pm 2.6 $ & G\\
7957.724 & 0.928 & $ -186.2 \pm 1.2 $ & $  268.2 \pm 2.1 $ & G\\
7957.730 & 0.931 & $ -186.4 \pm 1.2 $ & $  267.7 \pm 2.1 $ & G\\
7963.514 & 0.301 & $   80.8 \pm 0.6 $ & $ -199.2 \pm 2.3 $ & G\\
7963.518 & 0.302 & $   82.6 \pm 0.7 $ & $ -202.1 \pm 1.5 $ & G\\
7963.522 & 0.304 & $   84.3 \pm 0.7 $ & $ -204.8 \pm 1.9 $ & G\\
7963.526 & 0.305 & $   85.6 \pm 0.7 $ & $ -208.1 \pm 1.8 $ & G\\
7971.685 & 0.648 & $   45.3 \pm 0.5 $ & $ -109.9 \pm 1.2 $ & G\\
7971.691 & 0.651 & $   44.3 \pm 0.4 $ & $ -105.1 \pm 0.9 $ & G\\
7971.696 & 0.653 & $   42.8 \pm 0.4 $ & $ -100.4 \pm 0.8 $ & G\\
7971.702 & 0.655 & $   41.1 \pm 0.3 $ & $  -99.3 \pm 1.0 $ & G\\
7982.576 & 0.110 & $ -112.1 \pm 1.2 $ & $  149.7 \pm 0.8 $ & G\\
7982.580 & 0.112 & $ -110.9 \pm 1.3 $ & $  144.2 \pm 0.5 $ & G\\
7982.584 & 0.114 & $ -108.7 \pm 1.1 $ & $  142.7 \pm 1.1 $ & G\\
7982.588 & 0.115 & $ -107.5 \pm 1.2 $ & $  141.3 \pm 0.6 $ & G\\
7998.551 & 0.655 & $   40.5 \pm 0.5 $ & $  -96.2 \pm 1.3 $ & G\\
7998.556 & 0.657 & $   38.8 \pm 0.3 $ & $  -93.6 \pm 0.8 $ & G\\
7998.561 & 0.659 & $   37.4 \pm 0.2 $ & $  -89.6 \pm 1.1 $ & G\\
7998.567 & 0.662 & $   35.3 \pm 0.4 $ & $  -88.0 \pm 1.1 $ & G\\
8001.590 & 0.900 & $ -180.0 \pm 1.2 $ & $  254.4 \pm 1.5 $ & G\\
8001.596 & 0.903 & $ -179.9 \pm 1.1 $ & $  255.8 \pm 1.9 $ & G\\
8001.601 & 0.905 & $ -181.1 \pm 1.1 $ & $  258.9 \pm 2.4 $ & G\\
8001.607 & 0.907 & $ -181.7 \pm 1.1 $ & $  258.6 \pm 1.6 $ & G\\
8001.619 & 0.912 & $ -183.3 \pm 1.1 $ & $  259.4 \pm 1.6 $ & G\\
8001.624 & 0.914 & $ -183.4 \pm 1.3 $ & $  263.9 \pm 1.5 $ & G\\
8001.630 & 0.917 & $ -183.2 \pm 1.1 $ & $  263.8 \pm 2.2 $ & G\\
8001.635 & 0.919 & $ -183.4 \pm 1.1 $ & $  264.6 \pm 1.7 $ & G\\
8003.593 & 0.721 & $  -65.1 \pm 0.7 $ & $   32.3 \pm 1.3 $ & G\\
8003.599 & 0.724 & $  -59.9 \pm 2.2 $ & $    9.6 \pm 7.4 $ & G\\
8003.604 & 0.726 & $  -63.4 \pm 1.5 $ & $   17.9 \pm 4.5 $ & G\\
8003.610 & 0.728 & $  -61.8 \pm 2.4 $ & $    9.5 \pm 7.5 $ & G\\
8004.518 & 0.100 & $ -122.4 \pm 1.3 $ & $  160.9 \pm 0.8 $ & G\\
8004.523 & 0.102 & $ -120.5 \pm 1.3 $ & $  158.4 \pm 0.9 $ & G\\
8004.528 & 0.104 & $ -118.8 \pm 1.2 $ & $  155.4 \pm 1.0 $ & G\\
8004.534 & 0.107 & $ -116.3 \pm 1.2 $ & $  152.5 \pm 0.7 $ & G\\
8014.489 & 0.185 & $  -18.2 \pm 1.2 $ & $ -111.6 \pm 0.2 $ & G\\
8014.492 & 0.186 & $  -20.2 \pm 1.2 $ & $  -95.8 \pm 0.7 $ & G\\
8014.496 & 0.188 & $  -15.8 \pm 1.3 $ & $ -108.0 \pm 0.1 $ & G\\
8014.499 & 0.189 & $  -14.7 \pm 1.2 $ & $ -106.4 \pm 0.1 $ & G\\
8017.559 & 0.443 & $  151.6 \pm 0.9 $ & $ -332.1 \pm 1.1 $ & G\\
8017.565 & 0.445 & $  151.4 \pm 0.8 $ & $ -332.5 \pm 0.8 $ & G\\
8017.570 & 0.447 & $  151.3 \pm 0.8 $ & $ -331.7 \pm 1.2 $ & G\\
8017.576 & 0.450 & $  151.7 \pm 0.7 $ & $ -332.1 \pm 0.8 $ & G\\
8019.517 & 0.245 & $   40.1 \pm 1.0 $ & $  -97.7 \pm 0.4 $ & G\\
8019.522 & 0.247 & $   42.5 \pm 1.1 $ & $ -103.6 \pm 0.1 $ & G\\
8019.528 & 0.250 & $   44.6 \pm 1.0 $ & $ -109.6 \pm 0.1 $ & G\\
8019.533 & 0.252 & $   46.1 \pm 1.1 $ & $ -113.9 \pm 0.0 $ &G\\
8020.517 & 0.655 & $   38.0 \pm 0.5 $ & $  -98.2 \pm 0.8 $ & G\\
8020.523 & 0.657 & $   36.1 \pm 0.4 $ & $  -96.7 \pm 1.0 $ & G\\
8020.528 & 0.659 & $   34.6 \pm 0.3 $ & $  -93.7 \pm 1.0 $ & G\\
8020.534 & 0.662 & $   33.1 \pm 0.4 $ & $  -91.8 \pm 1.2 $ & G\\
8021.498 & 0.057 & $ -153.8 \pm 1.3 $ & $  210.1 \pm 1.1 $ &G\\
8021.503 & 0.059 & $ -152.1 \pm 1.2 $ & $  208.5 \pm 1.2 $ & G\\
8021.509 & 0.061 & $ -150.8 \pm 1.1 $ & $  206.5 \pm 1.5 $ &G\\
8021.514 & 0.063 & $ -149.1 \pm 1.2 $ & $  205.4 \pm 1.9 $ & G\\
\end{longtable}
\tablefoot{Column\,1 gives the heliocentric Julian date (HJD) of the observations at mid-exposure. Column\,2 gives the observational phase $\phi$ computed with the orbital period determined in Sect.\,\ref{sect:omegadot} (Table\,\ref{bestfitTable}). Columns\,3 and 4 give the radial velocities $RV_\text{P}$ and $RV_\text{S}$ of the primary and secondary stars, respectively. The errors represent $\pm1\sigma$. Column\,5 provides information about the instrumentation (F, H, and G stand for ESO 1.5 m + FEROS, ESO 3.6 m + HARPS, and ESO VLT + GIRAFFE, respectively).}}

\end{appendix}

\begin{thebibliography}{}
\bibitem[Asplund et al.(2009)]{Asplund}
Asplund, M., Grevesse, N., Sauval, A. J., \& Scott, P.\ 2009, \araa, 47, 481
\bibitem[Bailer-Jones et al.(2021)]{bailer21}
Bailer-Jones, C.~A.~L., Rybizki, J., Fouesneau, M., Demleitner, G., \& Andrae, R.\ 2021, \aj, 161, 147
\bibitem[Baroch et al.(2021)]{baroch21}
Baroch, D., Gim\'enez, A., Ribas, I., et al.\ 2021, \aap, 649, A64
\bibitem[Bouzid et al.(2005)]{bouzid05} 
Bouzid, M. Y., Sterken, C., \& Pribulla, T.\ 2005, \aap, 437, 769
\bibitem[Claret(1999)]{Claret99}
Claret, A.\ 1999, \aap, 350, 56
\bibitem[Claret et al.(2021)]{claret21}
Claret, A., Gim\'enez, A., Baroch, A., et al.\ 2021, \aap, 654, A17
\bibitem[Claret \& Torres(2019)]{claret19}
Claret, A., \& Torres, G.\ 2019, \apj, 876, 134
\bibitem[Conti \& Alschuler(1971)]{Conti71}
Conti, P. S., \& Alschuler, W. R.\ 1971, \apj, 170, 325
\bibitem[Gaia Collaboration(2021)]{EDR3}
Gaia Collaboration, Brown, A. G. A., Vallenari, A., Prusti, T., et al.\ 2021, \aap, 649, A1
\bibitem[Garc\'{\i}a \& Mermilliod(2001)]{Gar01}
Garc\'{\i}a, B., \& Mermilliod, J.-C.\ 2001, \aap, 368, 122
\bibitem[Gim\'enez \& Bastero(1995)]{gimenez95}
Gim\'enez, A., \& Bastero, M.\ 1995, \apss, 226, 99
\bibitem[Gonz\'alez \& Levato(2006)]{GL} 
Gonz\'alez, J. F., \& Levato, H.\ 2006, \aap, 448, 283
\bibitem[Gray(2008)]{Gray08}
Gray, D. F.\ 2008, The Observation and Analysis of Stellar Photospheres, 3rd edn.\ (Cambridge University Press)
\bibitem[Gray(2009)]{Gray}
Gray, R. O.\ 2009, A Digital Spectral Classification Atlas (\url{https://ned.ipac.caltech.edu/level5/Gray/frames.html})
\bibitem[Herrero et al.(1992)]{herrero92}
Herrero, A., Kudritzki, R. P., Vilchez, J. M., et al.\ 1992, \aap, 261, 209
\bibitem[Hill et al.(1974)]{Hil74}
Hill, G., Crawford, D. L., \& Barnes, J. V.\ 1974, \aj, 79, 1271
\bibitem[Hillier \& Miller(1998)]{Hillier}
Hillier, D. J., \& Miller, D. L.\ 1998, \apj, 496, 407
\bibitem[Hinkle et al.(2000)]{Hinkle}
Hinkle, K., Wallace, L., Valenti, J., \& Harmer, D.\ 2000, Visible and Near Infrared Atlas of the Arcturus Spectrum 3727-9300\,\AA, eds.\ K.\ Hinkle, L.\ Wallace, J.\ Valenti, \& D.\ Harmer, San Francisco: ASP
\bibitem[Humphreys \& McElroy(1984)]{humphreys} 
Humphreys, R. M., \& McElroy, D. B.\ 1984, \aj, 284, 565
\bibitem[Kaufer et al.(1999)]{Kaufer}
Kaufer, A., Stahl, O., Tubbesing, S., et al.\ 1999, The Messenger, 85, 8
\bibitem[Kuhn et al.(2017)]{Kuh17}
Kuhn, M. A., Getman, K. V., Feigelson, E. D., et al.\ 2017, \aj, 154, 214
\bibitem[Levato \& Malaroda(1980)]{levato80}
Levato, H., \& Malaroda, S.\ 1980, \pasp, 92, 323
\bibitem[Levato \& Morrell(1983)]{Lev83}
Levato, H., \& Morrell, N.\ 1983, Astrophys.\ Lett., 23, 183
\bibitem[Mahy et al.(2012)]{Mahy12}
Mahy, L., Gosset, E., Sana, H., et al.\ 2012, \aap, 540, A97
\bibitem[Martins(2011)]{Martins11}
Martins, F.\ 2011, Bulletin de la Soci\'et\'e Royale des Sciences de Li\`ege, 80, 29
\bibitem[Martins \& Hillier(2012)]{Martins12}
Martins, F., \& Hillier, D. J.\ 2012, \aap, 545, A95
\bibitem[Martins \& Plez(2006)]{MP}
Martins, F., \& Plez, B.\ 2006, A\&A, 457, 637
\bibitem[Mathys(1988)]{Mathys88}
Mathys, G.\ 1988, \aaps, 76, 427
\bibitem[Mayor et al.(2003)]{mayor03}
Mayor, M., Pepe, F., Queloz, D., et al.\ 2003, The Messenger, 114, 20 
\bibitem[Muijres et al.(2012)]{Muijres}
Muijres, L. E., Vink, J. S., de Koter, A., M\"uller, P. E., \& Langer, N.\ 2012, \aap, 537, A37
\bibitem[Palate \& Rauw(2012)]{Palate12}
Palate, M., \& Rauw, G.\ 2012, \aap, 537, A119
\bibitem[Pasquini et al.(2002)]{pasquini02} 
Pasquini, L., Avila, G., Blecha, A., et al.\ 2002, 110, 1
\bibitem[Perry et al.(1990)]{Per90}
Perry, C. L., Hill, G., Younger, P. F., \& Barnes, J. V.\ 1990, \aaps, 86, 415
\bibitem[Raucq et al.(2016)]{Raucq}
Raucq, F., Rauw, G., Gosset, E., et al.\ 2016, \aap, 588, A10 
\bibitem[Rauw et al.(2016)]{rauw16}
Rauw, G., Rosu, S., Noels, A., et al.\ 2016, \aap, 594, A33
\bibitem[Ricker et al.(2015)]{ricker15}
Ricker, G. R., Winn, J. N., Vanderspek, R., et al.\ 2015, JATIS 1, 014003
\bibitem[Rosu et al.(2020a)]{rosu20b}
Rosu, S., Noels, A., Dupret, M.-A., et al.\ 2020a, \aap, 642, A221
\bibitem[Rosu et al.(2020b)]{Rosu}
Rosu, S., Rauw, G., Conroy, K. E., et al.\ 2020b, \aap, 635, A145
\bibitem[Rosu et al.(2022)]{rosu22}
Rosu, S., Rauw, G., Farnir, M., Dupret, M.-A., \& Noels, A.\ 2022, \aap, 660, A120
\bibitem[Royer et al.(1998)]{Royer}
Royer, P., Vreux, J.-M., \& Manfroid, J.\ 1998, \aaps, 130, 407
\bibitem[Sana et al.(2005)]{sana05}
Sana, H., Antokhina, E., Royer, P., et al.\ 2005, \aap, 441, 213
\bibitem[Sana et al.(2008)]{San08}
Sana, H., Gosset, E., Naz\'e, Y., Rauw, G., \& Linder, N.\ 2008, \mnras, 386, 447
\bibitem[Sana et al.(2003)]{sana03}
Sana, H., Hensberge, H., Rauw, G., \& Gosset, E.\ 2003, \aap, 405, 1063
\bibitem[Sana et al.(2007)]{San07}
Sana, H., Rauw, G., Sung, H., Gosset, E., \& Vreux, J.-M.\ 2007, \mnras, 377, 945
\bibitem[Schmitt et al.(2016)]{schmitt16}
Schmitt, J. H. M. M., Schr\"oder, K.-P., Rauw, G., et al.\ 2016, \aap, 586, A104
\bibitem[Scuflaire et al.(2008)]{scuflaire08}
Scuflaire, R., Th\'eado, S., Montalb\'an, J., et al.\ 2008, \apss, 316, 83
\bibitem[Sim\'on-D\'{\i}az \& Herrero(2007)]{Simon-Diaz}
Sim\'on-D\'{\i}az, S., \& Herrero, A.\ 2007, \aap, 468, 1063
\bibitem[Slettebak et al.(1975)]{Slettebak}
Slettebak, A., Collins, G. W., Boyce, P. B., White, N. M., \& Parkinson, T. D.\ 1975, \apjs, 29, 137
\bibitem[Sota et al.(2014)]{Sota}
Sota, A., Ma\'{\i}z Apell\'aniz, J., Morrell, N. I., et al.\ 2014, \apjs, 211, 10
\bibitem[Sota et al.(2011)]{Sota11}
Sota, A., Ma\'{\i}z Apell\'aniz, J., Walborn, N. R., et al.\ 2011, \apjs, 193, 24
\bibitem[Sterken(1983)]{sterken83}
Sterken, C.\ 1983, The ESO Messenger, 33, 10
\bibitem[Sterken(1994)]{sterken94}
Sterken, C.\ 1994, in the Impact of Long-Term Monitoring on Variable-Star Research, NATO ARW, ed. C. Sterken, \& M. de Groot, NATO ASI Series C, 436, 1 (Kluwer AC. Publ.)
\bibitem[Struve(1944)]{struve44}
Struve, O.\ 1944, \apj, 100, 189
\bibitem[Sung et al.(1998)]{Sung98}
Sung, H., Bessel, M. S., \& Lee, S.-W.\ 1998, \aj, 115, 734
\bibitem[Sung et al.(2013)]{Sung13}
Sung, H., Sana, H., \& Bessell, M. S.\ 2013, \aj, 145, 37
\bibitem[Torres et al.(2010)]{torres10}
Torres, G., Andersen, J., \& Gim\'enez, A.\ 2010, \aapr, 18, 67
\bibitem[Voels et al.(1989)]{voels89}
Voels, S. A., Bohannan, B., Abbott, D. C., \& Hummer, D. G.\ 1989, \apj, 340, 1073
\bibitem[Walborn \& Fitzpatrick(1990)]{WP90}
Walborn, N. R., \& Fitzpatrick, E. L.\ 1990, \pasp, 102, 379
\bibitem[Wichmann(2011)]{Wichmann} 
Wichmann, R.\ 2011, Nightfall: Animated Views of Eclipsing Binary Stars, Astrophysics Source Code Library, record ascl:1106.016
\bibitem[Zacharias et al.(2013)]{Zacharias13}
Zacharias, N., Finch, C. T., Girard, T. M., et al.\ 2013, \aj, 145, 44
\end{thebibliography}
\end{document}